\documentclass[sigconf,screen]{acmart}

\renewcommand\footnotetextcopyrightpermission[1]{}

\AtBeginDocument{%
  \providecommand\BibTeX{{%
    \normalfont B\kern-0.5em{\scshape i\kern-0.25em b}\kern-0.8em\TeX}}}

\acmConference[ESEC/FSE 2020]{The 28th ACM Joint European Software Engineering Conference and Symposium on the Foundations of Software Engineering}{8 - 13 November, 2020}{Sacramento, California, United States}
\usepackage{graphicx}
\usepackage{enumitem}
\usepackage{subcaption}
\usepackage{amsmath}
\usepackage{xspace}
\usepackage{tcolorbox}

\theoremstyle{acmdefinition}
\newtheorem{exmp}{Example}[section]

\newcommand{\todo}[1]{\textcolor{black}{#1}}
\newcommand{\toolname}{\textsc{LibHarmo}\xspace}

\begin{document}

\title{Interactive, Effort-Aware Library Version Harmonization}

\author{Kaifeng Huang}
%\authornote{K. Huang, B. Chen, X. Peng, D. Zhou, Y. Wang and W. Zhao are with the School of Computer Science and Shanghai Key Laboratory of Data Science, Fudan University, China and the Shanghai Institute of Intelligent Electronics \& Systems, China.}
\affiliation{
\institution{Fudan University}
\country{China}
\vspace{-1pt}
}

\author{Bihuan Chen}
%\authornotemark[1]
%\authornote{B. Chen is the corresponding author.}
\affiliation{
\institution{Fudan University}
\country{China}
\vspace{-1pt}
}

\author{Bowen Shi}
%\authornotemark[1]
\affiliation{
\institution{Fudan University}
\country{China}
\vspace{-1pt}
}

\author{Ying Wang}
%\authornotemark[1]
\affiliation{
\institution{Fudan University}
\country{China}
\vspace{-1pt}
}

\author{Congying Xu}
%\authornotemark[1]
\affiliation{
\institution{Fudan University}
\country{China}
\vspace{-1pt}
}

\author{Xin Peng}
%\authornotemark[1]
\affiliation{
\institution{Fudan University}
\country{China}
\vspace{-1pt}
}

\begin{abstract}
% !TeX root = ../main.tex

As a mixed result of intensive dependency on third-party libraries,~flexible mechanism to declare dependencies, and increased number~of~modules in a project, multiple versions of the same third-party library~are directly depended in different modules of a project. Such library~version inconsistencies can increase dependency maintenance cost,~or even lead to dependency conflicts when modules are inter-dependent. Although automated build tools (e.g., Maven's \textit{enforcer} plugin)~provide partial support to detect library version inconsistencies, they~do not provide any support to harmonize inconsistent library versions.

We first conduct a survey with \todo{131} Java developers from GitHub to retrieve first-hand information about the root~causes, detection methods, reasons for fixing or not fixing, fixing strategies,~fixing efforts, and tool expectations on library version inconsistencies. Then, based on the insights from our survey, we propose~\toolname,~an~interactive, effort-aware library version harmonization technique,~to detect library version inconsistencies,~interactively suggest~a harmonized version with the least harmonization efforts based~on~library API usage~analysis, and refactor build configuration files. 

\toolname is currently developed for Java Maven projects. Our experimental~study on \todo{443} highly-starred Java Maven projects from GitHub indicates that i) \toolname identifies \todo{621} library version~inconsistencies covering \todo{152 (34.3\%)} of projects, and ii) the average harmonization efforts~are that \todo{1} and \todo{12} library API calls are affected, respectively due to the deleted and changed library APIs in the harmonized version. \todo{5} library version inconsistencies have~been~confirmed, and \todo{1} of them has been already harmonized by developers.

\end{abstract}

%%
%% The code below is generated by the tool at http://dl.acm.org/ccs.cfm.
%% Please copy and paste the code instead of the example below.
%%
%\begin{CCSXML}
%<ccs2012>
% <concept>
%  <concept_id>10010520.10010553.10010562</concept_id>
%  <concept_desc>Computer systems organization~Embedded systems</concept_desc>
%  <concept_significance>500</concept_significance>
% </concept>
% <concept>
%  <concept_id>10010520.10010575.10010755</concept_id>
%  <concept_desc>Computer systems organization~Redundancy</concept_desc>
%  <concept_significance>300</concept_significance>
% </concept>
% <concept>
%  <concept_id>10010520.10010553.10010554</concept_id>
%  <concept_desc>Computer systems organization~Robotics</concept_desc>
%  <concept_significance>100</concept_significance>
% </concept>
% <concept>
%  <concept_id>10003033.10003083.10003095</concept_id>
%  <concept_desc>Networks~Network reliability</concept_desc>
%  <concept_significance>100</concept_significance>
% </concept>
%</ccs2012>
%\end{CCSXML}
%
%\ccsdesc[500]{Computer systems organization~Embedded systems}
%\ccsdesc[300]{Computer systems organization~Redundancy}
%\ccsdesc{Computer systems organization~Robotics}
%\ccsdesc[100]{Networks~Network reliability}

%\keywords{}

\maketitle

% !TeX root = ../main.tex

\section{Introduction}\label{sec:intro}

With the increased diversity and complexity of modern systems,~modular development~\cite{Schlosser2004} has become a common practice to encourage reuse,~improve maintainability, and provide efficient ways for large teams of developers to collaborate~\cite{Humble2010}. Therefore, automated build tools (e.g., Maven) provide mechanisms (e.g., the \textit{aggregation} mechanism in Maven~\cite{projectdependency}) to support multi-module projects~for~the~ease~of management and build. In contrast to the benefits that multi-module project brings to software development, one of the drawbacks~is~the sophisticated dependency management (colloquially termed as ``dependency hell''~\cite{Jang2006}),  exacerbated by the increased number of modules and the intensive dependency on third-party libraries. In this paper, we focus on the dependency management in Maven projects as Maven has dominated the build tool market for many years~\cite{Paraschiv2018}.

% https://stackoverflow.com/questions/11730791/why-and-when-to-create-a-multi-module-maven-project

%Software artifacts are gaining more complexity and variety. Projects are organized in multiple modules to allow for building multiple artifacts with a shared release version so that those different could be used in different purposes and accessed separately. Reasons might be developers might want to deploy artifacts separately(e.g. build artifacts running on different platforms.), or consumers of the library might only want to use a subset of the class, or the developers might want a clear distinction between public APIs and private implementations. 

\textbf{Problem.} It is quite common that different modules of a project directly depend on the same third-party libraries. Maven provides flexible mechanisms for child modules to either inherit third-party~library dependencies from parent modules (e.g., the \textit{inheritance} mechanism~\cite{projectdependency}) or declare their own third-party library dependencies.~Besides, Maven allows the version of a third-party library dependency to be explicitly hard-coded or implicitly referenced~from a property which can be declared in parent modules. Therefore, \textit{library version inconsistency} can be easily caused in practice; i.e., multiple~versions of the same third-party library are directly depended~in different modules of a project. Even if the same version of a third-party~library is directly depended in different modules, the versions~can~be separately declared instead of referencing a common property.~We refer to it as \textit{library version false consistency} as it is likely to turn~into library version inconsistency when there is an incomplete library version update (e.g., a developer updates the version in one of the modules). Intuitively, library version inconsistency could increase dependency maintenance cost in the long run, or even~lead~to~dependency conflicts~\cite{Wang2018DCM} when modules are inter-dependent.

For example, an issue \texttt{HADOOP-6800}~\cite{HADOOP-6800} was reported to the~project \texttt{Apache} \texttt{Hadoop}, and said that ``\textit{multiple versions of the same library JAR are being pulled in .... Dependent subprojects use different versions. E.g. Common depends on Avro 1.3.2 while MapReduce depends~on~1.3.0. Since MapReduce depends on Common, this has the potential to cause a problem at runtime}''. This issue was prioritized as a blocker issue, and was resolved in 30 days. Developers found other library version inconsistencies, and finally harmonized the inconsistent versions of libraries \texttt{avro}, \texttt{commons-logging}, \texttt{commons-logging-api} and \texttt{jets3t} across modules \texttt{Common}, \texttt{MapReduce} and \texttt{HDFS}.

Maven's \textit{enforcer} plugin uses a \textit{dependency convergence} rule to detect multiple versions of the same third-party library along the~transitive dependency graph; i.e., if a module has two dependencies, \texttt{A} and \texttt{B}, and both depends on the same dependency, \texttt{C},~this~rule~will~fail the build if \texttt{A} depends on a different version of \texttt{C} than the version~of \texttt{C} depended~on by \texttt{B}. In that sense, this rule cannot detect library version inconsistencies across modules that are not inter-dependent, and does not provide any support to harmonize inconsistent library versions. As project developers have no direct~control~to~harmonize the~inconsistent library versions in transitive dependencies,~we~only consider direct dependencies across modules.

\textbf{Approach.} To better address the problem, e.g., by realizing~practical solutions that are acceptable by developers, it is important~to~first understand developers' practices on library version inconsistencies. Therefore, we conduct a survey with \todo{131} Java developers~from~GitHub to retrieve first-hand information about the root~causes, detection methods, reasons for fixing or not fixing, fixing strategies, fixing~efforts, and tool expectations on library version inconsistencies. \todo{90.8\%} of participants experienced library version inconsistency, and \todo{69.4\%} consider it as a problem in project maintenance. Our~survey suggests several insights, e.g., tools~are~needed to proactively~locate~and harmonize inconsistent library versions, and such tools need to interact with developers and provide API-level harmonization efforts.

Then, inspired by the insights from our developer survey,~we~propose \toolname, the first interactive, effort-aware technique~to~harmonize inconsistent library versions in Java Maven projects. \toolname works~in three steps. First, it identifies library version inconsistencies~by~analyzing build configuration files (i.e., POM files). Second, for each~library version inconsistency, it suggests a harmonized version with the least harmonization efforts (e.g., the number of calls to library APIs that are deleted and changed in the harmonized version) based on library~API~usage analysis and interaction with developers. Finally, if developers determine to harmonize, it refactors POM files, and also suggests replacement library APIs to some deleted library APIs based on API documentations.

We have run \toolname against \todo{443} highly-starred Java~Maven projects from GitHub. Our experimental results have indicated that i) \toolname detects \todo{621} library version inconsistencies, which~cover \todo{152 (34.3\%)} of projects, and ii) the average harmonization efforts~are that \todo{1} and \todo{2} of the \todo{24} called library APIs are respectively deleted and changed in the harmonized version, totally affecting \todo{1} and \todo{12} library API calls. Moreover, \todo{5} library version inconsistencies have~been confirmed, and \todo{1} of them has been harmonized by developers.

\textbf{Contributions.} This paper makes the following contributions.

\begin{itemize}[leftmargin=*]
\item We conducted the first survey with \todo{131} Java developers from~GitHub to retrieve first-hand information about the practices and tool~expectations on library version inconsistencies.

\item We proposed the first interactive, effort-aware library version~harmonization technique, \toolname, based on our survey insights.

\item We evaluated \toolname on \todo{443} highly-starred Java~Maven projects from GitHub, and found \todo{621} library version inconsistencies. \todo{5} of them have been confirmed with \todo{1} being harmonized.
\end{itemize}

% !TeX root = ../main.tex

\section{Developer Survey}\label{sec:survey}

Our online survey is designed for developers who participated~in~the development of Java Maven multi-module projects. Therefore,~we~selected Java Maven multi-module projects from GitHub, and also~restricted that the number of stars was larger than 200 to ensure~the project popularity. Finally, we had \todo{443} projects. From these projects, we collected \todo{5,316} developers whose email on profile page was valid. We sent an email to each of the \todo{5,316} developers to clarify the library version inconsistency problem and kindly ask them to participate~in our online questionnaire survey (the questions are shown in Table~\ref{table:questions}, and the complete questionnaire with options is available~at~\cite{website}).~We promised that their participation would remain confidential, and all the analysis and reporting would be based on aggregated responses.

Our survey consists of 14 questions, covering  the following seven aspects, to learn about their professional background,~practices and tool expectations on library version inconsistencies.

\begin{table}[!t]
    \centering
    \scriptsize
    \caption{Survey Questions}\label{table:questions}
    \vspace{-10pt}
    \begin{tabular}{|c|p{7.61cm}|}
    \hline
    Q1   & How many years of Java programming experience do you have?\\\hline
%    Q2   & Have you ever been participated in the development of multi-module Java project?\\\hline
    Q2   & How many modules in a Java project did you participate in?\\\hline
    Q3   & Have you ever encountered library version inconsistency?\\\hline
    Q4   & Is library version inconsistency a problem during project maintenance?\\\hline\hline
    Q5   & What are the root causes of library version inconsistencies?\\\hline
    Q6   & How did you detect library version inconsistencies?\\\hline
    Q7   & What are the reasons of not fixing library version inconsistencies?\\\hline
    Q8   & What are the reasons of fixing library version inconsistencies?\\\hline
    Q9  & Which version do you use as the harmonized version to fix library version inconsistencies?\\\hline
    Q10   & How do you fix library version inconsistencies?\\\hline
    Q11   & How much time do you spend in fixing library version inconsistencies?\\\hline
    Q12   & Which part of it is most time-consuming in fixing library version inconsistencies?\\\hline\hline
    Q13   & Is an automatic library version harmonization tool useful for library management?\\\hline
    Q14   & Which features would be useful for an automatic library version harmonization tool?\\\hline
    \end{tabular}
\end{table}

\textbf{Professional Background (Q1--Q4).} In response to the invitation emails, \todo{131} developers finished the questionnaire within~seven days (i.e., a participation rate of \todo{2.5\%}). Of all participants, \todo{44.3\%}~have more than 10 years of Java programming experience, \todo{25.2\%} have~5~to 10 years, and \todo{30.5\%} have less than 5 years. \todo{47.3\%} participated~in~the development of more than 10 modules in one  project, \todo{23.7\%} participated in 5 to 10 modules, and \todo{29.0\%} participated in less than~5~modules. \todo{90.8\%} of participants experienced library version~inconsistency, and \todo{69.4\%} consider it as a problem in project maintenance. The participants have relatively good experience in modular development as well as in handling library version inconsistencies.

\textbf{Root Causes (Q5).} \todo{67.1\%} and \todo{65.8\%} named unawareness of the same library in other modules and backward incompatibility issues in library versions as the major root causes of library version~inconsistencies. Different development schedule among different~modules (\todo{46.1\%}), unawareness of the library version inconsistency~problem (\todo{31.6\%}), and not regarding library version inconsistency as a problem (\todo{23.7\%}) are the further root causes. Other minor root causes (\todo{14.5\%}) include bad dependency management hygiene, unawareness of new library versions, usage difficulty with Maven, etc.

\textbf{Detection Methods (Q6).} Being asked about the detection or manifestation of library version inconsistencies, bugs due to conflicting library versions~\cite{Wang2018DCM} (\todo{72.4\%}) is the main way to manifest, followed by bugs due to library API behavior changes (\todo{47.4\%}). Manual investigation of module POM files (\todo{46.1\%}) is the main way~to~detect, followed by communication with developers~of~other~modules (\todo{14.5\%}) and adoption of Maven's \textit{enforcer} plugin (\todo{10.5\%}).

\textbf{Reasons for Fixing or not Fixing (Q7--Q8).} The participants reported four main reasons for not fixing: heavy fixing efforts due to backward incompatibility issues (\todo{45.3\%}), heavy fixing efforts due to intensive library API dependency (\todo{38.7\%}), fixing difficulty~due~to~different development schedule in different modules (\todo{36.0\%}), and no serious consequence occurred (\todo{30.7\%}). \todo{6.6\%} emphasized that they always selected to fix. On the other hand, there are three main reasons for fixing: avoiding great maintenance efforts in the long run (\todo{68.4\%}), ensuring consistent library API behaviors  across modules (\todo{63.2\%}), and serious consequences occurred (e.g., bugs)~(\todo{55.3\%}).

\textbf{Fixing Strategies (Q9--Q10).} When harmonizing the inconsistent library versions, \todo{77.6\%} used one of the newer versions than all currently declared versions with the least harmonization efforts,~but \todo{29.0\%} chose one of the currently declared versions with the least harmonization efforts. Besides, \todo{61.8\%} harmonized the versions in all of the affected modules, while \todo{38.2\%} only harmonized the versions in some of the affected modules.

\textbf{Fixing Efforts (Q11--Q12).} \todo{50.0\%} spent hours in fixing library version inconsistencies, \todo{32.9\%} even spent days, and only \todo{11.8\%}~spent minutes. Besides, locating all inconsistent library versions (\todo{56.7\%}), determining the harmonized version (\todo{49.3\%}), and refactoring the source code (\todo{48.0\%}) are the most time-consuming steps in fixing. Other time-consuming steps include refactoring the POM~files~(\todo{32.0\%}) and verifying the fix through regression testing (\todo{6.7\%}).

\textbf{Tool Expectations (Q13--Q14).} \todo{45.6\%} thought an automated~library version harmonization tool would be useful, but \todo{14.0\%} thought it would not be useful mostly because they already adopted Maven's \textit{enforcer} plugin. \todo{46.5\%} thought it depended on how well it would~be integrated into the build process, how automated it would be, etc. With respect to the most useful feature in such a tool, detecting all library version inconsistencies (\todo{75.9\%}) and suggesting the harmonized version (\todo{71.4\%}) are the most useful ones, followed by reporting detailed API-level fixing efforts (\todo{49.1\%}) and refactoring the POM files (\todo{42.0\%}). Surprising, refactoring the source code (\todo{25.0\%}) is less useful than all the previous features. 

\begin{figure*}[!t]
    \centering
    \includegraphics[scale=0.35]{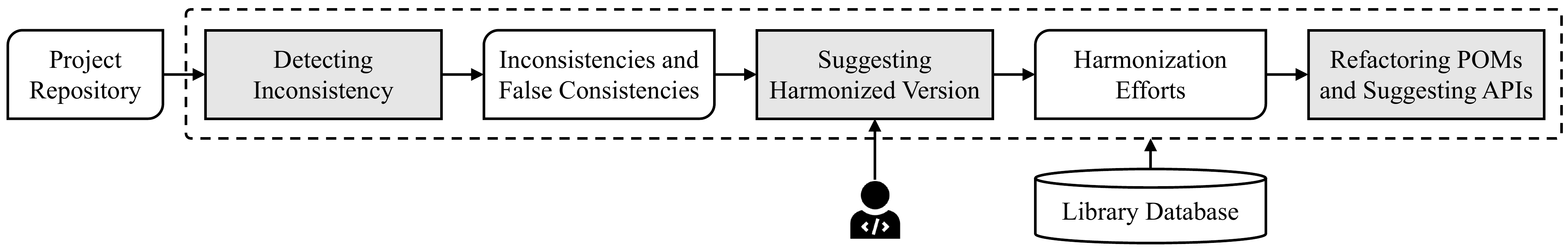}
    \vspace{-5pt}
    \caption{An Overview of \toolname}
    \label{fig:overview} 
\end{figure*}

\textbf{Insights.} From our survey results, we have several insights. \textbf{I1:} tools are needed to help developers proactively locate and harmonize inconsistent library versions, as library version~inconsistencies are mostly manually detected, or passively found~after serious~consequences. \textbf{I2:} developers should interact with such tools to determine where and whether to harmonize,~as~library version~inconsistencies span multiple modules that have different development schedule, and might be not fixed due to heavy harmonization efforts. \textbf{I3:} such tools need to provide developers~with API-level harmonization~efforts, as API backward~incompatibility, API dependency intensity, and API behavior consistency are key factors for developers to determine whether to harmonize. \textbf{I4:} such tools need to be integrated into the build~process for the ease of adoption.

% !TeX root = ../main.tex

\section{Methodology}\label{sec:method}

Based on the insights \textbf{I1}, \textbf{I2} and \textbf{I3} from our developer survey,~we~propose the first interactive, effort-aware technique, named \toolname,~to assist developers in harmonizing inconsistent library versions (and falsely consistent library versions). As shown in Fig.~\ref{fig:overview}, it takes~as~an input a Java Maven project repository, and interactively works with developers in three steps, i.e., detecting inconsistency~(Sec.~\ref{subsec:detect}),~suggesting harmonized~version (Sec.~\ref{subsec:suggest}), and refactoring POMs and suggesting~APIs~(Sec.~\ref{subsec:refactor}). \toolname also relies on a library database (Sec.~\ref{subsec:database}) to provide JAR files and documentations. Currently, \toolname is at the stage of a prototype, and thus it is not integrated into the build process and does not satisfy the insight \textbf{I4}.

\subsection{Detecting Inconsistency}\label{subsec:detect}

The first step of \toolname is composed of three sub-steps:~it~first~generates the POM inheritance graph, then analyzes the inheritance~relations to resolve library dependencies in each POM, and finally~identifies library version inconsistencies and false consistencies.

\textbf{Generating POM Inheritance Graph.} Maven provides the \textit{inheritance} mechanism~\cite{projectdependency} to inherit elements (e.g., dependency) from a parent POM. It does not support multiple inheritance,~however,~it indirectly supports the concept  by using the \textit{import} scope~\cite{importscope}.~Maven also does not allow cyclic inheritance. Therefore, the inheritance relations among POMs in a project form a directed acyclic graph.~We define such a POM inheritance graph $\mathcal{G}$ as a 2-tuple $\langle \mathcal{M},\mathcal{E}\rangle$, where~$\mathcal{M}$ denotes all the POMs in a project, and $\mathcal{E}$ denotes the inheritance relations among the POMs in $\mathcal{M}$. Each inheritance relation $e \in \mathcal{E}$ is denoted as a 2-tuple $\langle m_1, m_2\rangle$, where $m_1, m_2 \in \mathcal{M}$, and $m_1$~inherits $m_2$ (i.e., $m_2$ is the parent POM of $m_1$).

To construct $\mathcal{G}$ of a project, \toolname scans its repository recursively to collect all the local POMs and put them into $\mathcal{M}$. Then, for each POM $m$ in $\mathcal{M}$, \toolname parses it to locate its parent POMs based on the \textit{inheritance} mechanism and the \textit{import} scope;~i.e.,~\toolname parses the \texttt{parent} section and the \texttt{dependencyManagement} section. For each located parent POM $m'$, an inheritance relation $e = \langle m, m'\rangle$ is generated and put into $\mathcal{E}$. As $m'$ can be a remote~POM, \toolname crawls it from Maven repository, and puts it into $\mathcal{M}$. $\mathcal{E}$~is constructed after all the local and remote POMs in $\mathcal{M}$ are parsed.

%Notice that $\mathcal{G}$ can be a set of graphs.

\begin{figure}
    \centering
    \includegraphics[scale=0.32]{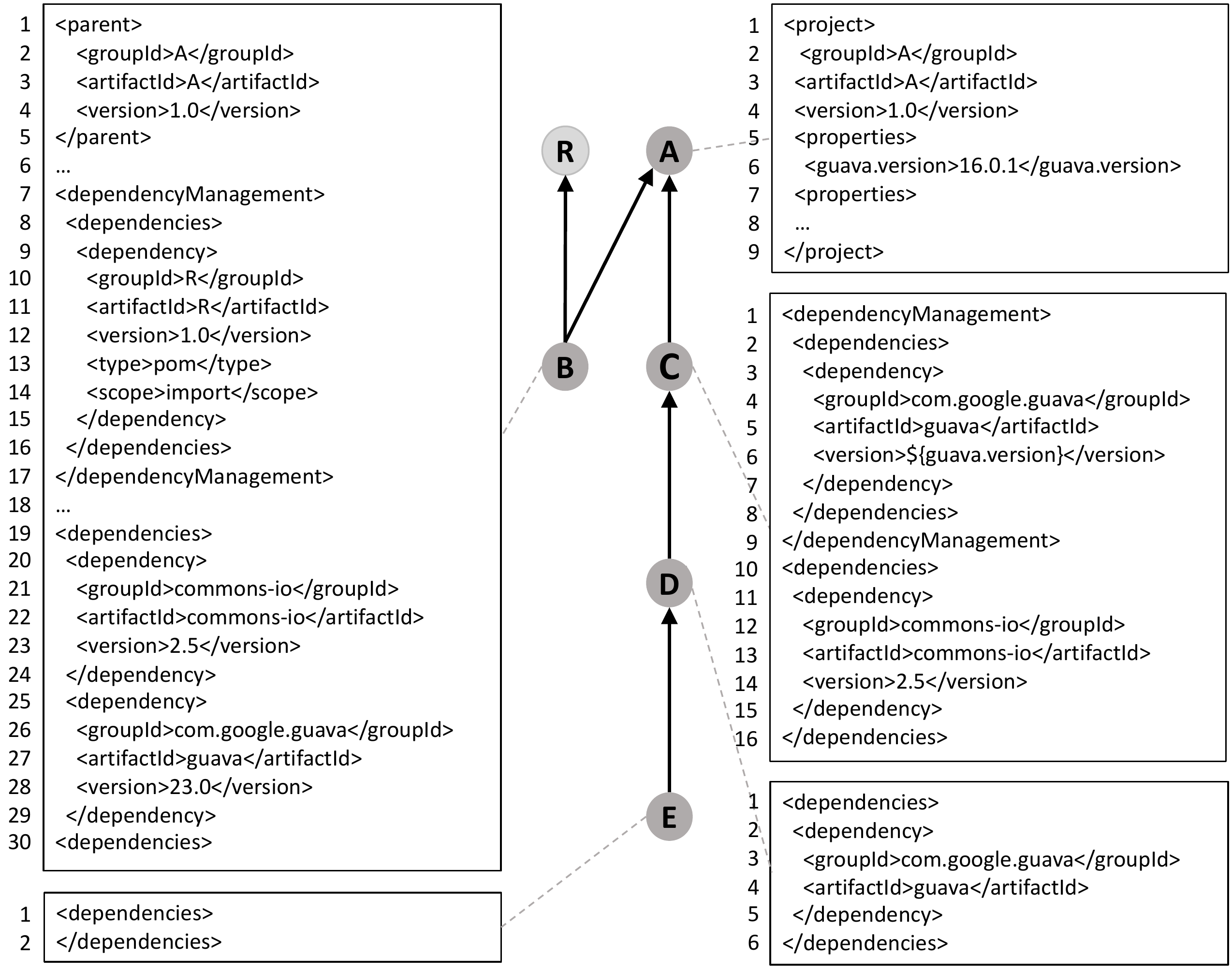}
    \vspace{-5pt}
    \caption{An Example of POM Inheritance Graph}
    \label{fig:example} 
\end{figure}

\begin{exmp}
Fig.~\ref{fig:example} presents a generated POM~inheritance graph, where the nodes represent POMs, the arrows represent inheritance relations, and the dotted lines link to excerpts from POMs. Here~\texttt{A},~\texttt{B}, \texttt{C}, \texttt{D} and \texttt{E} are local POMs, and \texttt{R}~is~a~remote POM. \texttt{B} has two parent POMs, \texttt{A} and \texttt{R}. In particular, \texttt{B} inherits \texttt{A} by declaring the \texttt{groudId}, \texttt{artifactId} and \texttt{version} of \texttt{A} in the \texttt{parent} section (Line 1--5~in~\texttt{B}). \texttt{B} inherits \texttt{R} by declaring the \texttt{groudId}, \texttt{artifactId} and \texttt{version} of \texttt{R} in a dependency with \texttt{type} being \textit{pom} and \texttt{scope} being \textit{import} in the \texttt{dependencyManagement} section (Line 7--17 in \texttt{B}).
\end{exmp}

% \textbf{Identification of Modular Project.} A multi-module project follows by the rule that's given by maven pom specifications. It is built from an aggregator POM that manages a group of submodules. Mostly, the aggregator is located in the project's root directory. Now, the submodules are regular Maven projects, and they can be built separately or through the aggregator POM. Through a inheritance parsing of the aggregator POM, we can get a collection of all the descendant POMs. On the other hand, those dissoiate POMs will not be part of the Maven build process, thus will be ignored in project build. And those with no aggregator POM is recognized as non multi-module project. We follow the build rationale of maven. That is, given a project \textit{p}, we start searching from the aggregator POM, namely the root POM in directory, and obtain a set of POM files $\mathcal{M}$ that are declared recursively as the modules of the parent POM. Also, we denote $\mathcal{E}$ as the edges of the POM file which represents the inheritance relationships of $\mathcal{M}$. Thus, an inheritance graph $\mathcal{G} = \langle \mathcal{M},\mathcal{E}\rangle$ is generated. For files in projects, we denotes as $\mathcal{F} = \{f_1,f_2,f_3,\cdots\}$ and for every file $f$, $f_{module}$ denotes the module it belongs to and $f_{module} \in \mathcal{M}$. We will refer to each POM as a module in the latter context.

\textbf{Resolving Library Dependencies.}  We first introduce Maven's dependency declaration mechanisms before diving into the details. The~\texttt{dependencies} section contains the library dependencies~that~a POM declares to use, and such library dependencies will be automatically inherited by child POMs, whereas~the~\texttt{dependencyManagement} section contains the library dependencies~that~a POM declares to manage, and such library dependencies will be used/inherited only when they are explicitly declared in the \texttt{dependencies} section~without specifying their version. Moreover, the version~of~a~library~dependency can be explicitly declared by a hard-coded value or implicitly declared via referencing a property. A property can~be~overwritten by declaring the same property with a different value.

\begin{exmp}
In Fig.~\ref{fig:example}, \texttt{B} declares two library dependencies~\texttt{B}~wants to use, and the versions are hard-coded (Line 20--29 in \texttt{B}). \texttt{C} declares one library dependency \texttt{C} wants to use (Line 10--16 in \texttt{C}); and \texttt{C} also declares one library dependency \texttt{C} wants to manage (Line 1--9~in~\texttt{C}), and the version references a property, \texttt{guava.version},~which~is declared in Line 5--7 in \texttt{A}. \texttt{D} automatically inherits the library dependency in Line 10--16 in \texttt{C}; and \texttt{D} also inherits the managed~library dependency in Line 1--9~in~\texttt{C} by explicitly declaring it in Line 1--6~in \texttt{D}. \texttt{E} inherits from \texttt{D} the two library dependencies \texttt{D} inherits from \texttt{C}.
\end{exmp}

Based on the dependency declaration mechanisms,~all~the~library dependencies of a POM can be resolved based on the resolved~library dependencies of its ancestor POMs. To ease~the~detection~and harmonization of inconsistencies and false consistencies, we first define a library dependency $d$ as a 6-tuple~$\langle lib, ver, pro, m_{lib}, m_{ver}, m_{pro}\rangle$, where $lib$ denotes a library, uniquely identified by its \texttt{groupId}~(i.e., the organization $lib$ belongs to) and \texttt{artifactId} (i.e., the name~of~$lib$); $ver$ denotes the resolved version number~of~$lib$;~$pro$ denotes~the~property that the version of $lib$ references, and~it~will be \texttt{null} when~the~version of $lib$~is~hard-coded; $m_{lib}$~denotes the POM that owns $lib$ either by~declaration~or inheritance; $m_{ver}$ denotes the POM that declares the version of $lib$; and $m_{pro}$~denotes the~POM that declares $pro$, and it will be \texttt{null} when $pro$ is \texttt{null}. 

For each POM $m$~in~$\mathcal{M}$, we resolve $m$' library dependencies $\mathcal{D}_m$ that are declared~in~$m$~or~inherited from ancestors of $m$. To this end, \toolname~performs a breath-first search on $\mathcal{G}$ to visit~$m$~and~$m$'s ancestors while following Maven's ``nearest definition wins'' and~``first declaration wins'' strategy~\cite{importscope}. For each~visited POM, we parse~each~library dependency~in the \texttt{dependencies} section to create~a~$d$~and~put $d$ to $\mathcal{D}_m$,~and~analyze the \texttt{properties} and \texttt{dependencyManagement} section to resolve the unresolved version of library dependencies~in $\mathcal{D}_m$. Finally, we get all library dependencies $\mathcal{D} = \bigcup_{m\in\mathcal{M}} \mathcal{D}_m$.

%There could be multiple versions of library to choose from due to inherited library version and version declared in its own file. And theres is also a mechanism called \textit{nearest definition}, which means the nearest definition to the target POM file wins.

%First, for each $m$ in $\mathcal{M}$, \toolname~ first parses the target POM file, gets all libraries that is declared on its own. Second, \toolname~ searches with a breath first strategy on $\mathcal{G}$ to get all the silent libraries to the target POM file. Lastly, \toolname~ resolve the versions of all the libraries it gets by following the rule of \textit{nearest definition}. If version is declared implicitly, \toolname~ finds a version declaration on $\mathcal{G}$ also with a breath first strategy.  

% \toolname~ adopts a depth first search to iterate all descendant modules while keeping track of all the properties it declares to resolve dependencies in modules. Although properties declared in parent module is visible to all its sub-modules. The previously defined ones could be overridden in sub-modules if it is delcared again with a different version value. So we leverages a stack-based variable referencing dictionary while doing a DFS search, which pops in all the properties declares before that module, and pops out when finished. Then multiple declarations occurs, we chose a declaration near to the top. 

\begin{table}[!t]
\footnotesize
\centering
\caption{An Example of Resolving Library Dependencies}\label{table:resolve}
\vspace{-10pt}
\begin{tabular}{|c|l|l|}
\hline
E & & \\\hline
D & <guava, \xspace\xspace\xspace\xspace\xspace\xspace\xspace\xspace\xspace\xspace, \xspace\xspace\xspace\xspace\xspace\xspace\xspace\xspace\xspace\xspace\xspace\xspace\xspace\xspace\xspace\xspace\xspace\xspace\xspace\xspace\xspace\xspace\xspace\xspace, E, \xspace\xspace\xspace, \xspace\xspace\xspace> & \\\hline
C & <guava, \xspace\xspace\xspace\xspace\xspace\xspace\xspace\xspace\xspace\xspace, guava.version, E, C, \xspace\xspace\xspace> & <commons-io, 2.5, null, E, C, null> \\\hline
A & <guava, 16.0.1, guava.version, E, C, A> & <commons-io, 2.5, null, E, C, null> \\\hline
\end{tabular}
\end{table}

\begin{exmp}
Table~\ref{table:resolve} presents the the process of resolving library dependencies for \texttt{E} in Fig.~\ref{fig:example} along its inheritance hierarchy.~At~\texttt{E},~as \texttt{E} does not declare any library dependency, no library dependency~is created. Next, at \texttt{E}'s parent \texttt{D}, \texttt{guava} is declared but its version~is~not declared. Hence, $d_1$ is created with $lib$ and $m_{lib}$ set to \texttt{guava}~and~\texttt{E}. Next, at \texttt{D}'s parent \texttt{C}, $d_1$'s version is declared~by~referencing~a~property. Thus, $d_1$'s $pro$ and $m_{ver}$ is set~to \texttt{guava.version} and \texttt{C}. Meanwhile, \texttt{C} declares \texttt{commons-io} and hard-codes its version. Thus,~$d_2$~is created as $\langle$\texttt{commons-io}, \texttt{2.5}, \texttt{null}, \texttt{E}, \texttt{C}, \texttt{null}$\rangle$. Finally, at \texttt{C}'s parent \texttt{A}, the property \texttt{guava.version} is declared, and thus $d_1$'s $ver$ and $m_{pro}$ is set to \texttt{16.0.1}  and  \texttt{A}.
\end{exmp}

%Project aggregation is designed to let parent POM (project) know its modules, so that when a maven build command is invoked in parent directory, it will also be executed in modules as well. By following maven's build rule, those have modules specified in parent project have to specify in each POM who their parent POM is.

%By means of \textit{project project aggregation}, we can obtain a set of POM files as the connected module files rooted from a super POM file. Here we refer to those as module POM files. Our assumption is that, inconsistency issues happens within those connected modules. And those separately organized POMs, which is not part of a build process from root projects, is not regarded as module POM files. Thus we filter out those POMs.

\textbf{Identifying Inconsistencies and False Consistencies.} As we do not have direct control over remote POMs, we remove from~$\mathcal{D}$~the library dependencies whose $m_{lib}$ is a remote POM. However, ~it~is~possible that the library dependencies of local POMs are inherited from remote POMs. To detect library version inconsistencies and~false~consistencies, we first identify the libraries $\mathcal{L}$ from $\mathcal{D}$, i.e., $\mathcal{L} = \{d.lib \mid d \in \mathcal{D}\}$. Then, for each $lib \in \mathcal{L}$, we find all the library dependencies $\mathcal{D}_{lib} = \{d \mid d \in \mathcal{D} \land d.lib = lib\}$. Finally, we determine~the~consistency of $\mathcal{D}_{lib}$  by classifying it into the following four types.

\begin{itemize}[leftmargin=*]
\item \textit{Inconsistency (IC).} $\mathcal{D}_{lib}$ belongs to the type of \textit{inconsistency} if the library dependencies in $\mathcal{D}_{lib}$ do not have the same version; i.e., $\mathcal{D}_{lib}$ satisfies $|\mathcal{D}_{lib}| > 1 \land \exists d_1, d_2 \in \mathcal{D}_{lib}, d_1.ver \neq d_2.ver$.

\item \textit{True Consistency (TC).} $\mathcal{D}_{lib}$ belongs to the type of \textit{true consistency} if all the library dependencies in $\mathcal{D}_{lib}$ have the same version~by referencing~one property; i.e., $\mathcal{D}_{lib}$ satisfies $|\mathcal{D}_{lib}| > 1 \land \forall d_1, d_2 $ $\in \mathcal{D}_{lib}, d_1.pro \neq null \land d_1.pro = d_2.pro \land d_1.m_{pro} = d_2.m_{pro}$.
    
\item \textit{False Consistency (FC).} $\mathcal{D}_{lib}$ belongs to the type of \textit{false consistency} if all the library dependencies in $\mathcal{D}_{lib}$ have the same version but do not reference one property (i.e., the version is resolved by referencing different properties or by hard-coding); i.e., $\mathcal{D}_{lib}$ satisfies $\exists d_1, d_2 \in \mathcal{D}_{lib}, d_1.pro \neq null \land d_2.pro \neq null \land (d_1.pro \neq d_2.pro \lor d_1.m_{pro} \neq d_2.m_{pro}) \lor \exists d \in \mathcal{D}_{lib}, d.pro = null$.

\item \textit{Single Library (SL).} $\mathcal{D}_{lib}$ belongs to the type  of \textit{single library} if there is only one library dependency in $\mathcal{D}_{lib}$ (i.e., $|\mathcal{D}_{lib}| = 1$). 
\end{itemize}

\begin{figure}
    \centering
    \includegraphics[scale=0.4]{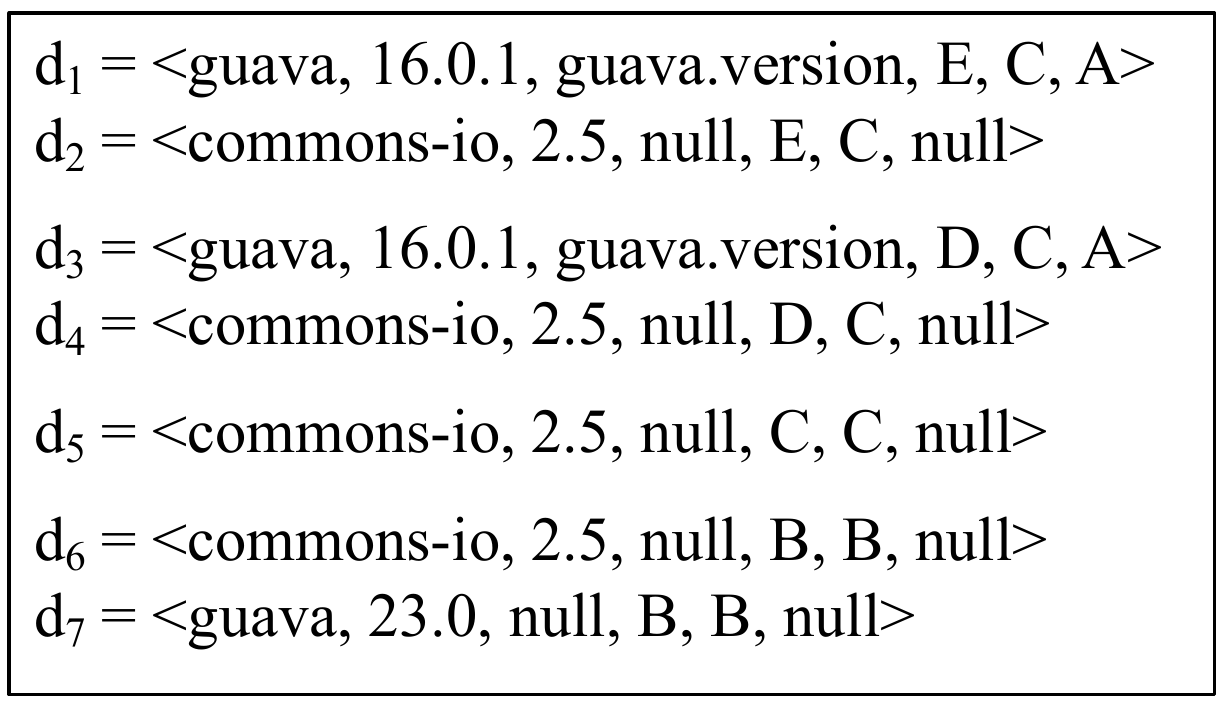}
    \vspace{-5pt}
    \caption{The Resolved Library Dependencies of Fig.~\ref{fig:example}}
    \label{fig:exampleD} 
\end{figure}

\begin{exmp}\label{example:d}
Fig.~\ref{fig:exampleD} presents all the resolved library dependencies of Fig.~\ref{fig:example}, which involve two libraries \texttt{guava} and \texttt{commons-io}. Hence, we have $\mathcal{D}_{guava} =\{d_1, d_3, d_7\}$ and $\mathcal{D}_{commons-io} =\{d_2, d_4, d_5, d_6\}$. $\mathcal{D}_{guava}$ belongs to \textit{IC}, and $\mathcal{D}_{commons-io}$ belongs to \textit{FC}.
\end{exmp}

% \textbf{Strategy 1. Mild Update}. The mild update should be appropriate for those who are not familiar and try to maintain what the project is. The goal is to update library versions while remain what the declarations are. The process should be mild and affect the least module properties structure. It mainly collects the information about library versions and update to a common one. 

% \textbf{Strategy 2. Erase Hardcoding}.Hardcoding is time-
% consuming when doing maintenance jobs as explained in previous section. Thus we brought up a strategy called hardcoding-eraing, which is to group explicit versions into one properties and keep other properties still while update to one as many as possible. The main advantage of the techiniques is two-fold. First it erases harcoding, helping improve efficiency. And second the existing declared property will not be affected by a new declared property.

% \textbf{Strategy 3. Avoid confusion of Properties}. Multiple declared properties are also annoying and confusing when doing update jobs. This strategy aims to unify implicit versions variables into one properties while keeping explicit versions declarations still. 

\subsection{Suggesting Harmonized Version}\label{subsec:suggest}

%For files in projects, we denotes as $\mathcal{F} = \{f_1,f_2,f_3,\cdots\}$ and for every file $f$, it belongs to a POM project $\mathcal{M}$.

For a false consistency, the same version is adopted across different library dependencies, and it is likely to turn into an inconsistency~if there is an incomplete library version update (e.g., a developer only updates the version of some of the library dependencies).~Hence,~it also needs to be harmonized to become a true consistency (which will be introduced  in Sec.~\ref{subsec:refactor}). Here we~directly~suggest the currently used version as the harmonized~version to reduce~the~harmonization efforts. On the other hand,~for~an inconsistency $\mathcal{D}_{lib}$,~we~first~analyze the harmonization efforts at the library API level, and then~interactively suggest a harmonized version with the least efforts.

\textbf{Analyzing Harmonization Efforts.} Basically, we measure~the harmonization efforts in terms of the number of calls to library~APIs that are deleted or changed in the harmonized version, because the deleted library APIs may cause program crashes, while the changed library APIs may introduce API breaking. Hence, for each $d \in \mathcal{D}_{lib}$, \toolname~first applies JavaParser~\cite{smith2017javaparser} on the \texttt{src} folder that has the same prefix path to $d.m_{lib}$, together with the JAR files from~our library database (see Sec.~\ref{subsec:database}), to locate API calls to $d$. Thus,~we~have a set of called library APIs $\mathcal{A}_d$ and a set of library API calls $\mathcal{C}_d$.

Then, \toolname determines the candidate library versions $\mathcal{V}_d$~for harmonization from our library database which contains all the~released versions of $d.lib$. Here, we compute $\mathcal{V}_d$ as the versions that are no older than the highest version in $\mathcal{D}_{lib}$ as developers tend~to use newer versions, as suggested by our survey. Next, for each~candidate version $v$, \toolname locates the called library APIs in $d$ that are deleted or changed in the candidate version $v$. Here, an library API is deleted in $v$ if there is no library API with the~same~fully~qualified name in $v$. An library API is changed in $v$ if its fully qualified name is not changed but the body code of the library API~or~the~code of the methods in its static call graph is changed. \toolname uses java-callgraph~\cite{callgraph} to extract the static call graph. Thus, we decompose $\mathcal{A}_d$ into three sets $\mathcal{AD}_d^v$, $\mathcal{AC}_d^v$ and $\mathcal{AU}_d^v$, respectively~representing the called library APIs in $d$ that are deleted, changed and unchanged in $v$. Correspondingly, we can decompose $\mathcal{C}_d$ into three sets $\mathcal{CD}_d^v$, $\mathcal{CC}_d^v$ and $\mathcal{CU}_d^v$, respectively~representing the calls~to~the library APIs in $\mathcal{AD}_d^v$, $\mathcal{AC}_d^v$ and $\mathcal{AU}_d^v$ (i.e., the calls to the deleted, changed and unchanged library APIs). Therefore, the efforts~$f_d^v$~to harmonize $d$ to the version $v$ can be characterized as a 6-tuple, i.e., $f_d^v = \langle \mathcal{AD}_d^v, \mathcal{AC}_d^v, \mathcal{AU}_d^v, \mathcal{CD}_d^v, \mathcal{CC}_d^v, \mathcal{CU}_d^v \rangle$.

\textbf{Interactively Recommending Harmonized Version.} As revealed by our survey (see Sec.~\ref{sec:survey}), developers may choose~to~not~harmonize all inconsistent library dependencies due to various reasons (e.g., different development schedule, or heavy efforts due to~API~dependency intensity or backward incompatibility). Thus,~\toolname is designed to interact with developers such that 1) developers are  provided with detailed library API-level harmonization efforts $f_d^v$ for each library dependency $d \in \mathcal{D}_{lib}$ to be harmonized into each candidate version $v \in \mathcal{V}_d$; 2) developers have the flexibility~to~decide which of the library dependencies~$\mathcal{D}_{lib}' \subseteq \mathcal{D}_{lib}$~need~to~be~harmonized; and 3) developers are provided with a ranked list of candidate versions based on flexible combinations of $\mathcal{AD}_d^v$, $\mathcal{AC}_d^v$, $\mathcal{AU}_d^v$, $\mathcal{CD}_d^v$, $\mathcal{CC}_d^v$ and $\mathcal{CU}_d^v$ (e.g., the default ranking~is~based~on~the~summation of $|\mathcal{CD}_d^v|$ and $|\mathcal{CC}_d^v|$ over all library dependencies in $\mathcal{D}_{lib}'$) such that they can choose the harmonized version $v_h$ with the least harmonization efforts they consider acceptable.

To ease the determination of $\mathcal{D}_{lib}'$, we first decompose $\mathcal{D}_{lib}$~according to $d.m_{ver}$; i.e., the library dependencies that have their version declared in the same POM $m_{ver}$ are grouped into $\mathcal{D}_{lib}^{m_{ver}}$.~$\mathcal{D}_{lib}^{m_{ver}}$ actually belongs to the type of true consistency (or single library), and should be harmonized together to still keep the consistency.~For example, $\mathcal{D}_{guava}$ in Example~\ref{example:d} can be decomposed into $\mathcal{D}_{guava}^C = \{d_1, d_3\}$ and $\mathcal{D}_{guava}^B = \{d_7\}$. Based on the decomposition, we allow developers to determine which groups need~to~be harmonized.

%For each inconsistent library, $\mathcal{D}_{ic}^{\prime}$, there will be several library candidates to choose $Can_{i} = \{i|i =1,2,\cdots,n \}$. Each version of library dependency have to migrate to a suggested one. Thus there will be a migration stability set denoted as $S_{Can_{i}} = \{i|i=1,2,\cdots,n\}$, in which we choose one with most migration stability. The migration stability for each candidate will be the sum of stability in all dependencies which is $S_{Can_{i}} = \sum_{j=1}^{m}S^{j}_{Can_{i}}$, which $S^{j}_{Can_{i}}$ denotes a single migration stability for inconsistent library $l_j$ to candidate $l_i$. 

%We use $API_{deleted}$, $API_{same}$ and $API_{modified}$ to term API difference where $API_{deleted}$ denotes APIs that could not be found in candidate version. And $API_{same}$ denotes one API that all of its callsites, including method signature and method body remains the same. If there exists one changed method signature or method body, it will be labeled as $API_{modified}$. Also for those modified APIs, we have $Method_{modified}$, $Method_{deleted}$ and $Method_{added}$ in call graph pairs.

%$S^{j}_{can_{i}} = \mid API_{same} \mid+ \sum{1-\frac{Method_{added} + Method_{deleted}+Method_{modified}}{Method_{all}}}$.

\subsection{Refactoring POMs and Suggesting APIs}\label{subsec:refactor}

The last step of \toolname is to provide support to carry out~the~harmonization on POMs and source code. \toolname can automatically refactor POMs based on the library dependencies~$\mathcal{D}_{lib}'$~that~developers choose from an inconsistency $\mathcal{D}_{lib}$ and the harmonized~version $v_h$ that developers choose. The POM refactoring is exactly the same for false consistencies. Besides, \toolname provides conservative support for library API adaptation; i.e., it only suggests replacement library APIs to some of the deleted library APIs based~on~the~extracted information from API documentations.

\textbf{Refactoring POMs.} The goal of our harmonization is to make $\mathcal{D}_{lib}'$ become a true consistency; i.e., all the library dependencies~in $\mathcal{D}_{lib}'$ need to have their version reference a property of value $v_h$.~To this end, \toolname first locates the POMs $\mathcal{M}'$ that declare the~version of the library dependencies in $\mathcal{D}_{lib}'$; i.e.,  $\mathcal{M}' = \{d.m_{ver} \mid d \in \mathcal{D}_{lib}'\}$. On one hand, the lowest common ancestor of the POMs in $\mathcal{M}'$ on the POM inheritance graph $\mathcal{G}$ is the POM where \toolname newly declares a property of value $v_h$. On the other hand, $\mathcal{M}'$ contains the POMs where \toolname changes the (implicit or explicit) version declaration of $lib$ to a reference to the newly~declared~property. Occasionally, the lowest common ancestor could be a remote POM that we do not have direct control, or $\mathcal{G}$ contains~several~sub-graphs that are not connected. Thus,~\toolname~finds~several~lowest common ancestors, each of which~is~the~lowest~common ancestor~of some POMs in $\mathcal{M}'$, and then applies the similar refactoring process.

% finds a common and nearest local parent to all the declaring versions modules to declare a variable name, which we refers as a common nearest parent POM $m_{c}$ where a variable is visible in all its child POM files. 
 
% The algorithm of finding a common nearest parent POM is pretty easy. Given a list of $m_{declareversion}$, \toolname first picks up one POM file as an anchor file, the the rest of $m_{declareversion}$ is taken one by one to find a common parent POM file along the inheritance graph from bottom up. It navigates on the $\mathcal{G}$ while searching a first common parent with a POM file pair from bottom up. And anchor file updates to the common parent before the next round begins. And finally the anchor file will be the common parent POM file for all the $m_{declareversion}$.

%Given migrated version, $m_{declareversion}$ and $m_{c}$, we use \textit{maven-model} to load and parse the original POM file, get dependency node of the library, edit its version declaration into a variable reference, and then write back to file. Similarly, the common nearest parent POM file is loaded and added by a property declared with the migrated version value.

Finally, \toolname checks whether the properties that are referenced in $\mathcal{D}_{lib}'$ are referenced by the other library dependencies in $\mathcal{D}$. Specifically, for each library dependency $d \in \mathcal{D}_{lib}'$ that declares the version by referencing a property, \toolname~extracts a 2-tuple $\langle d.pro, d.m_{pro}\rangle$, and checks whether there exists a library dependency $d'$ in $\mathcal{D} - \mathcal{D}_{lib}'$ such that $d.pro = d'.pro \land d.m_{pro} = d'.m_{pro}$. If exists, $d.pro$ is still referenced by other library dependencies, and thus it is kept; otherwise, \toolname deletes $d.pro$ from $d.m_{pro}$.

\begin{figure}
    \centering
    \includegraphics[scale=0.31]{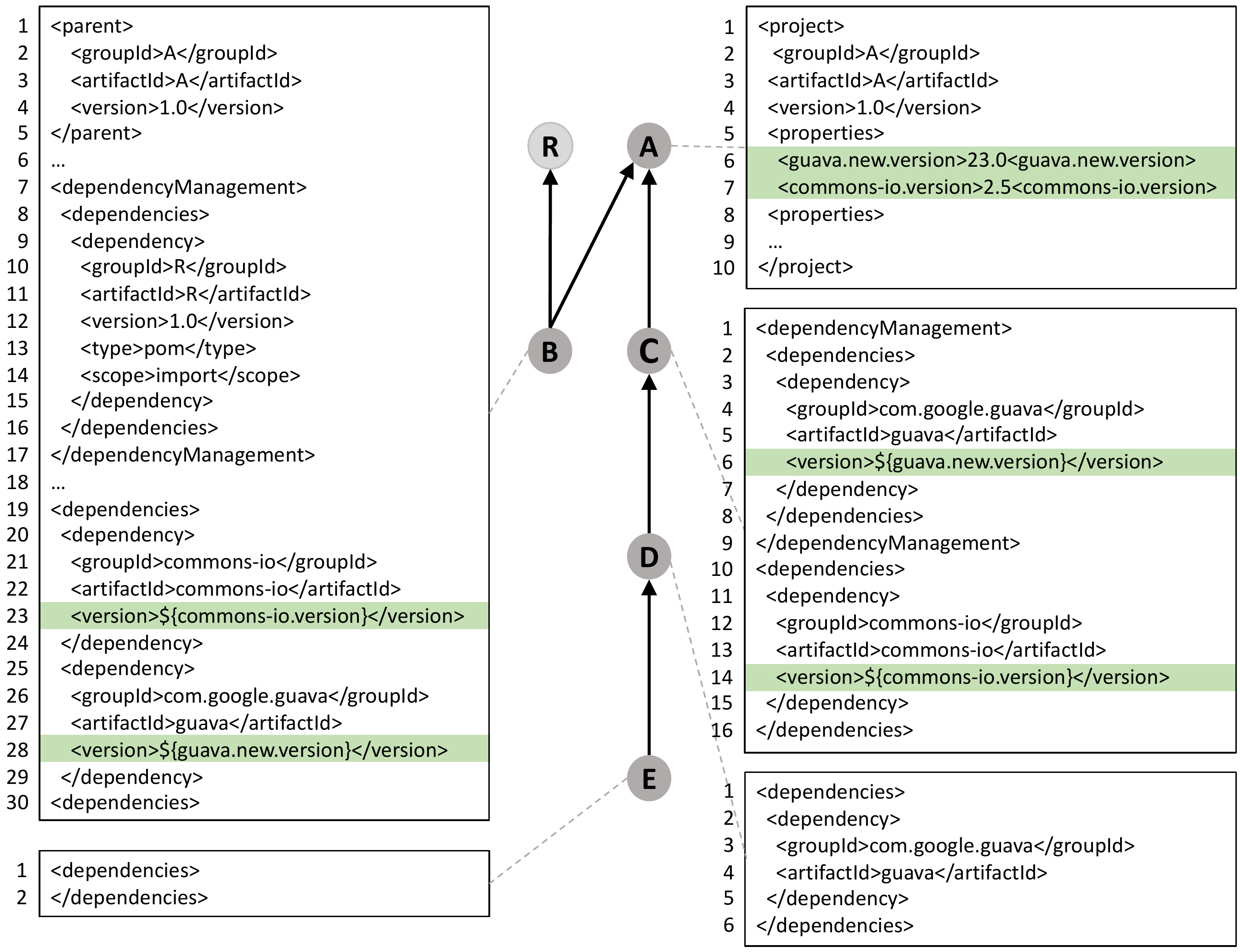}
    \vspace{-15pt}
    \caption{An Example of Refactoring POMs}
    \label{fig:exampleH} 
\end{figure}

\begin{exmp}
Given $\mathcal{D}_{guava} =\{d_1, d_3, d_7\}$ in Example~\ref{example:d} and~the harmonized version \texttt{23.0}, $\mathcal{M}'$ is computed as $\{\texttt{B}, \texttt{C}\}$, and their~lowest common ancestor is \texttt{A}. Hence, a property \texttt{guava.new.version}~is declared at Line 6 in \texttt{A} in Fig.~\ref{fig:exampleH}, and \texttt{B} and \texttt{C} respectively change~the version declaration at Line 28 in \texttt{B} and at Line 6 in \texttt{C} to reference~the property \texttt{guava.new.version}. Moreover, as the previous property \texttt{guava.version} is not referenced by other library dependencies, it is deleted from \texttt{A}. Similarly, for the false consistency $\mathcal{D}_{commons-io}$ $=\{d_2, d_4, d_5, d_6\}$ in Example~\ref{example:d}, $\mathcal{M}'$ is computed as $\{\texttt{B}, \texttt{C}\}$. Hence,  a property \texttt{commons-io.version} is declared in \texttt{A}, the lowest common ancestor of \texttt{B} and \texttt{C}; and the version declaration at Line 23 in \texttt{B} and at Line 14 in \texttt{C} is changed to reference \texttt{commons-io.version}.
\end{exmp}

\textbf{Suggesting APIs.} After POMs are  refactored, the source code also needs to be adapted to the harmonized version. Library API~adaptation has been widely investigated~\cite{Chow1996SUA,Balaban2005RSC,henkel2005catchup,xing2007api,dagenais2009semdiff,dagenais2011recommending,schafer2008mining,Nguyen2010GAA,fazzini2019automated,wu2010aura}, and empirical~studies~\cite{cossette2012seeking, wu2015impact} have shown that they achieved~an average accuracy of 20\%. Besides, our survey indicates that refactoring the source code is surprisingly a less useful feature. Based~on~the two observations, \toolname takes a conservation strategy to only consider the library APIs that are deleted in the harmonized version and attempt to suggest their replacement library APIs. The reason~is that some library APIs are deleted because they are deprecated and their replacement library APIs might be clearly documented~in~the \textit{deprecated} page in Javadoc in the form of ``use \texttt{xxx}'' where \texttt{xxx}~is the fully qualified name of a replacement library API. 

Specifically, for each deleted library API $a \in \mathcal{AD}_d^v$, \toolname first obtain from our library database the Javadocs of all the library releases released between the release date of the library version $d.ver$ and the release data of the harmonized version $v$. Notice that these library releases could possibly document the deprecation~of~$a$. Then, for each Javadoc, \toolname checks in the \textit{deprecated} page whether~the~library API $a$ is deprecated; and if yes, \toolname uses pattern matching to search the existence of ``use \texttt{xxx}''. If exists, the replacement library API to $a$ is found, and suggested to developers.

\subsection{Library Database}\label{subsec:database}

Recall that our harmonization efforts analysis (see Sec.~\ref{subsec:suggest}) requests from the library database the JAR files of a library version and some newer releases of the same library, and our replacement library~API suggestion (see Sec.~\ref{subsec:refactor}) requests the Javadocs of some~library releases from the library database. Therefore, \toolname crawls the JAR files and Javadocs of all releases of a library in a demand-driven way from Maven repository. Besides, \toolname regularly updates any new library releases for the libraries in our library database.

\section{Evaluation}\label{sec:evaluation}

We have implemented a prototype of \toolname in Java and Python in a total of \todo{14.6}K lines of code. \todo{We have released the source code~at our website~\cite{website}}. In this section, we report~our experimental results of \toolname on GitHub projects.

\begin{figure}[!t]
    \centering
    \begin{subfigure}[b]{0.23\textwidth}
    \centering
    \includegraphics[width=0.99\textwidth]{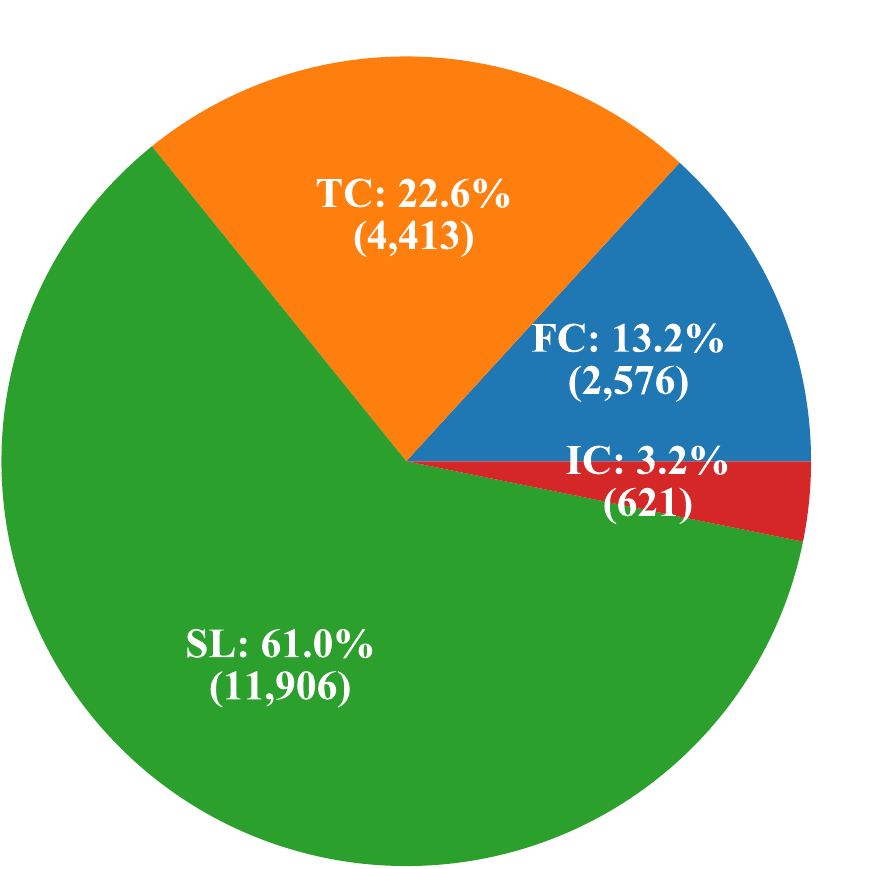}
    \caption{Distribution}
    \label{fig:rq1-distribuion}
    \end{subfigure}
    ~
    \begin{subfigure}[b]{0.23\textwidth}
    \centering
    \includegraphics[width=0.99\textwidth]{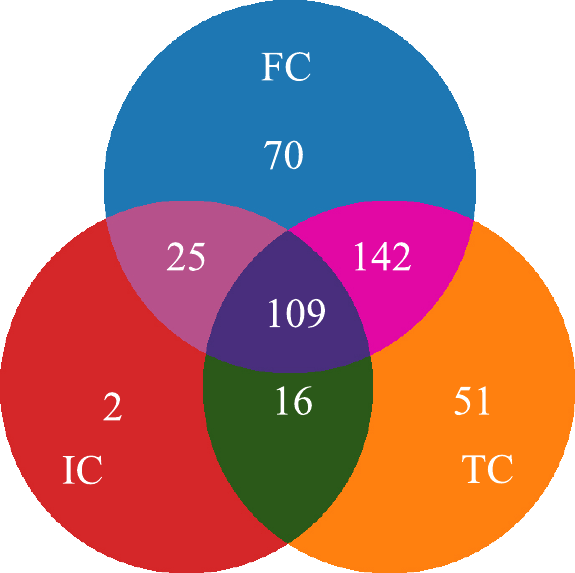}
    \caption{Intersection}
    \label{fig:rq1-intersection}
    \end{subfigure}
    \vspace{-5pt}
    \caption{Overall Distribution of IC, FC, TC and SL}
\end{figure}

 \begin{figure}[!t]
    \centering
    \begin{subfigure}[b]{0.45\textwidth}
    \centering
    \includegraphics[width=0.99\textwidth]{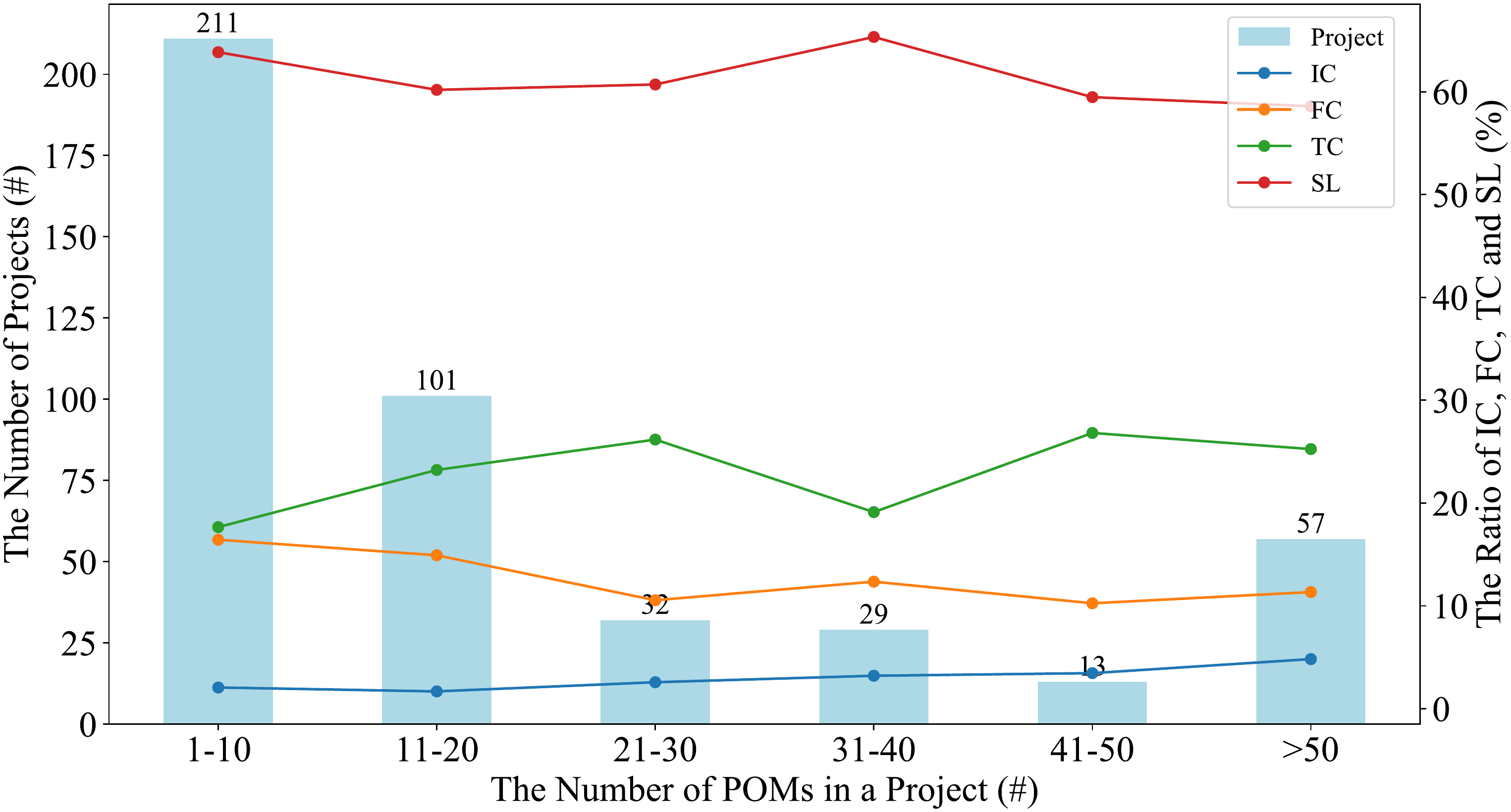}
    \vspace{-5pt}
    \caption{Distribution}
    \label{fig:rq1-distribution-poms}
    \end{subfigure}
    \begin{subfigure}[b]{0.45\textwidth}
    \centering
    \includegraphics[width=0.99\textwidth]{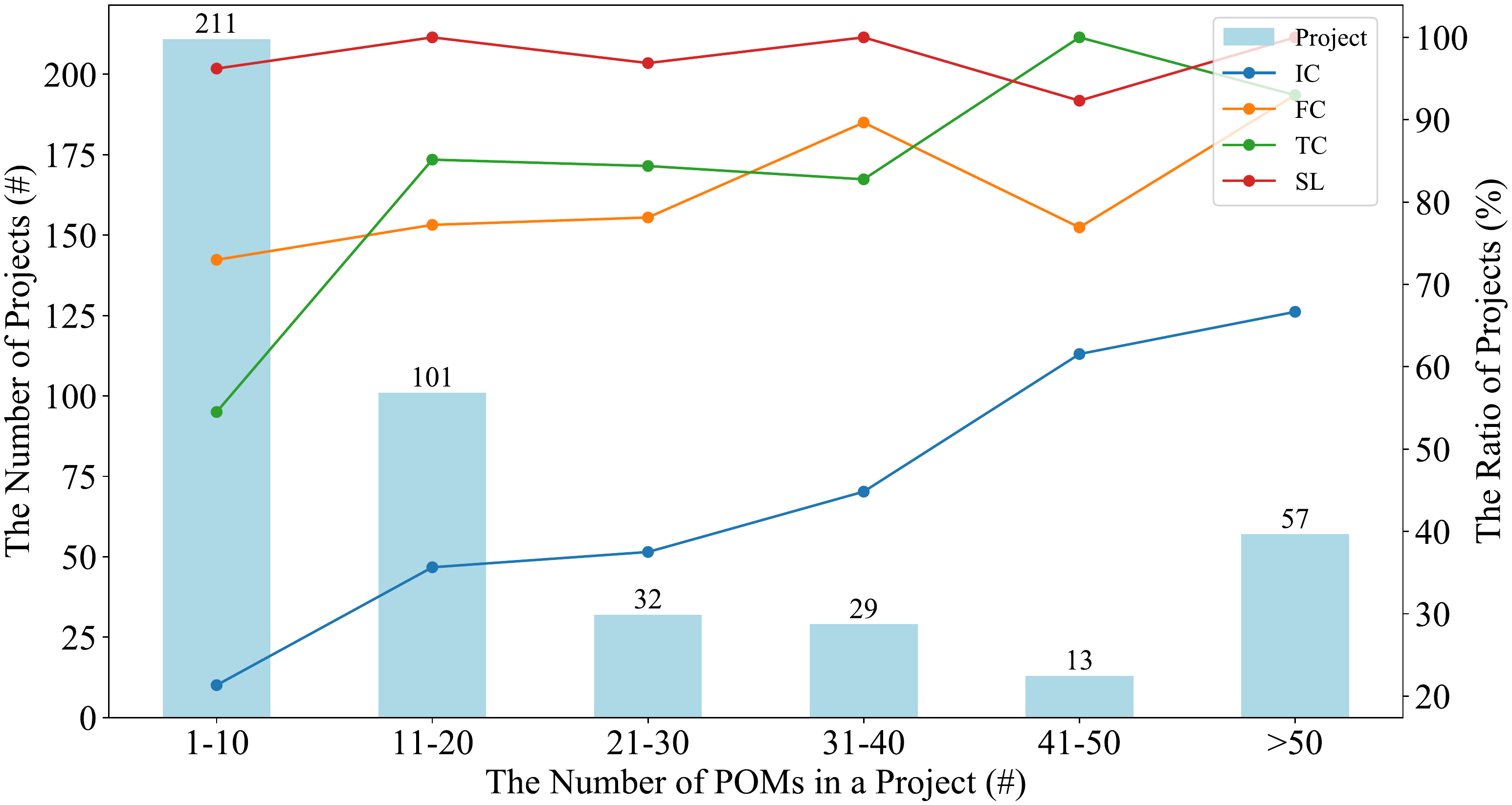}
    \vspace{-5pt}
    \caption{Projects Having IC, FC, TC and SL}
    \label{fig:rq1-distribution-poms-affect}
    \end{subfigure}
    \vspace{-10pt}
    \caption{Fine-Grained Distribution of IC, FC, TC and SL}
\end{figure}

\subsection{Evaluation Design}

%Our study aims to get an overview of inconsistency among multi-module projects. To evaluate \toolname~, we collected 918 java projects from Github to study the pervasiveness of inconsistency problem. The selection of projects is based on the popularity-number of stars. Also the projects should be a maven projects. Of these projects, we select projects with multiple modules declared in one of their POM files, which restricted our selection to 

We used the same set of \todo{443} Java Maven multi-module projects~used in our survey as the dataset for our evaluation. We designed~our~evaluation to answer the following four research questions.

\begin{itemize}[leftmargin=*]
\item \textbf{RQ1:} What is the distribution of inconsistency, false consistency, true consistency, and single library? (Sec.~\ref{subsec:distribution})

\item \textbf{RQ2:} What is the severity of the detected inconsistency~and~false consistency? (Sec.~\ref{subsec:severity})

\item \textbf{RQ3:} What are the efforts for harmonizing inconsistency? (Sec.~\ref{subsec:effort})

\item \textbf{RQ4:} What is  developers' feedback about \toolname? (Sec.~\ref{subsec:feedback})
\end{itemize}

Specifically, we ran \toolname against each project to 1) detect~all the inconsistencies and false consistencies, which is used~to~answer \textbf{RQ1} and \textbf{RQ2}, 2) analyze the~harmonization efforts for each inconsistency for each candidate harmonized version, which~is~used~to~answer~\textbf{RQ3}, and 3) generate a report including the previous two set~of information and send it to developers, which is used to answer \textbf{RQ4}.

\subsection{Distribution Evaluation (RQ1)}\label{subsec:distribution}

We first measured the overall distribution of inconsistency, false~consistency, true consistency and single library, and then measured~their fine-grained distribution with respect to the modular complexity~of projects (approximated as the number of POMs).

\textbf{Overall Distribution.} Fig.~\ref{fig:rq1-distribuion} shows the overall distribution of inconsistency (IC), false consistency (FC), true consistency (TC), and single library (SL) (see Sec.~\ref{subsec:detect}). SL accounts for \todo{61.0\%}, and IC, FC and TC account for \todo{39.0\%}, which means that more than one-third of the libraries are used across multiple modules. More specifically, TC accounts for \todo{22.6\%}, which is much higher than IC and FC, and covers \todo{318} projects. This indicates that library version harmonization (via referencing a property) is already a practice~that~is~adopted~by~many projects. Nevertheless, there are still~\todo{2,576}~cases of FC. They account for \todo{13.2\%} and cover \todo{346} projects. They are very likely to turn into~IC if not carefully maintained, and thus increase the burden of library maintenance. There are \todo{621} cases of IC, which~account~for~\todo{3.2\%}~and cover \todo{152} projects. These results~indicate that library version inconsistency and false consistency are common in real-world projects.

Moreover, Fig.~\ref{fig:rq1-intersection} reports the intersections among the projects that  are affected by IC, FC and TC. Noticeably, there is a high~overlap (i.e., \todo{251} projects) between TC and FC. This indicates that~while~many projects adopt consistent library versions, they still leave many libraries not truly consistent. Similarly, the libraries in \todo{51} projects~are all consistent, while the libraries in \todo{70} projects are all falsely consistent. Moreover, the overlap between FC and IC is also high (i.e., \todo{134} projects), and most of the projects that have IC also have FC.~This~is potentially because that FC has a high chance to turn into~IC. Furthermore, \todo{109} projects have IC, FC and TC at the same time, which indicates that using consistent library versions is not consistently recognized across the whole development team of a project.

\begin{figure*}[!t]
    \centering
    \begin{subfigure}[b]{0.465\textwidth}
    \centering
    \includegraphics[width=0.99\textwidth]{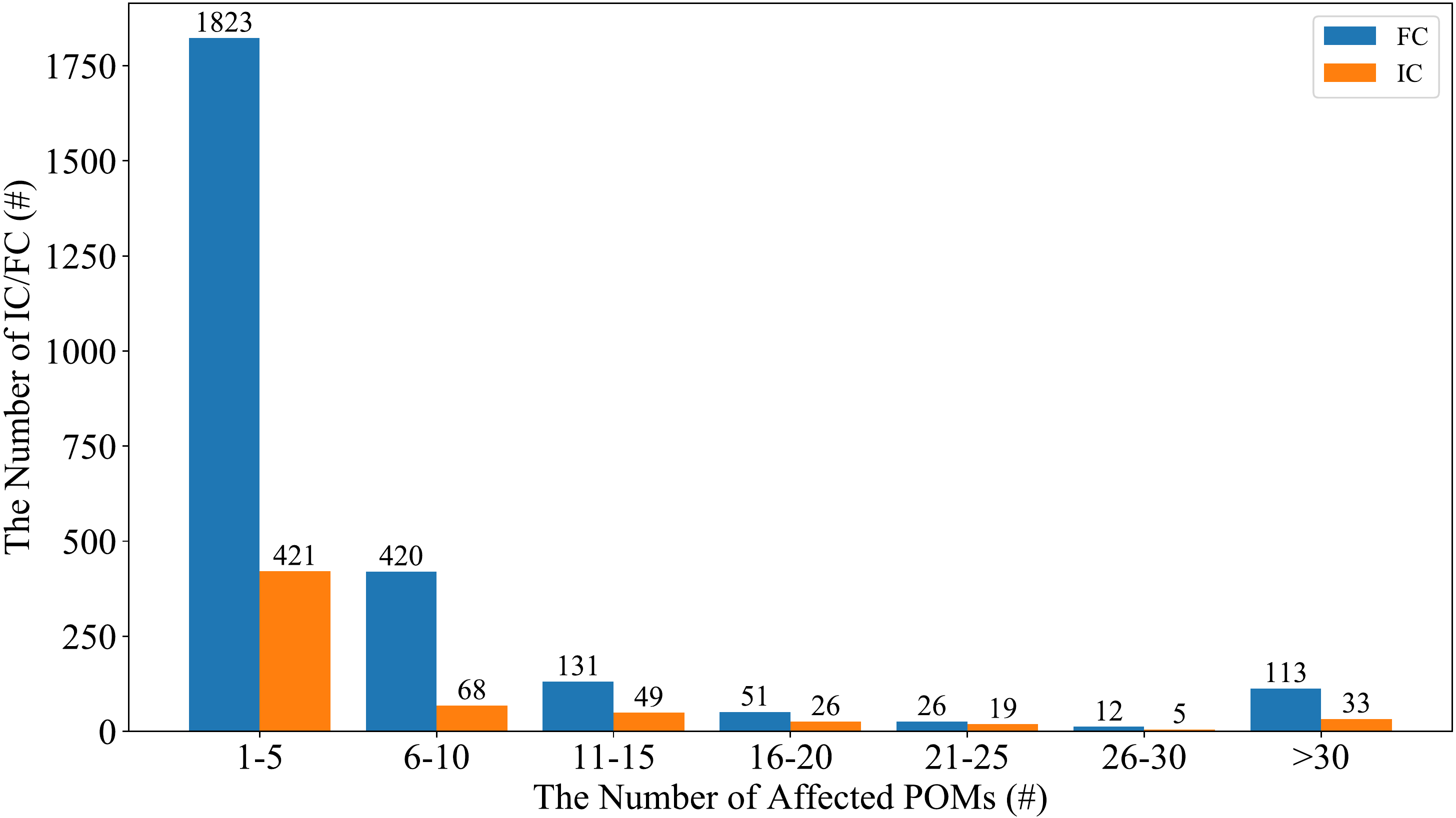}
    \vspace{0pt}
    \caption{Distribution across the Number of Affected POMs}
    \label{fig:rq2-abs}
    \end{subfigure}
    ~~
    \begin{subfigure}[b]{0.46\textwidth}
    \centering
    \includegraphics[width=0.99\textwidth]{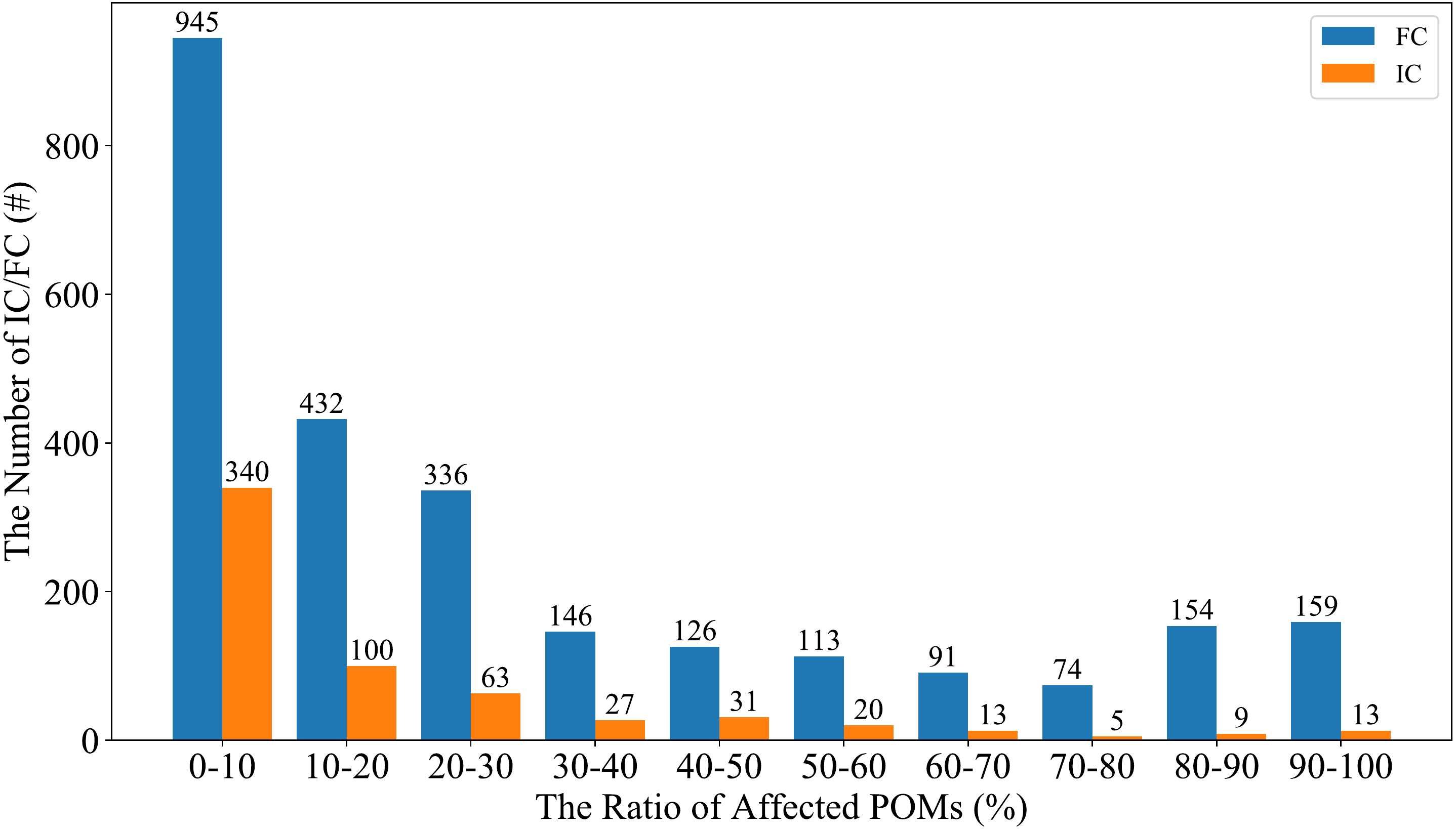}
    \vspace{0pt}
    \caption{Distribution across the Ratio of Affected POMs}
    \label{fig:rq2-rel}
    \end{subfigure}
    \vspace{-10pt}
    \caption{Distribution of IC and FC across Affected POMs}
    \label{fig:rq2-affected} 
\end{figure*}

\begin{figure*}[!t]
    \centering
    \begin{subfigure}[b]{0.43\textwidth}
    \centering
    \includegraphics[width=0.99\textwidth]{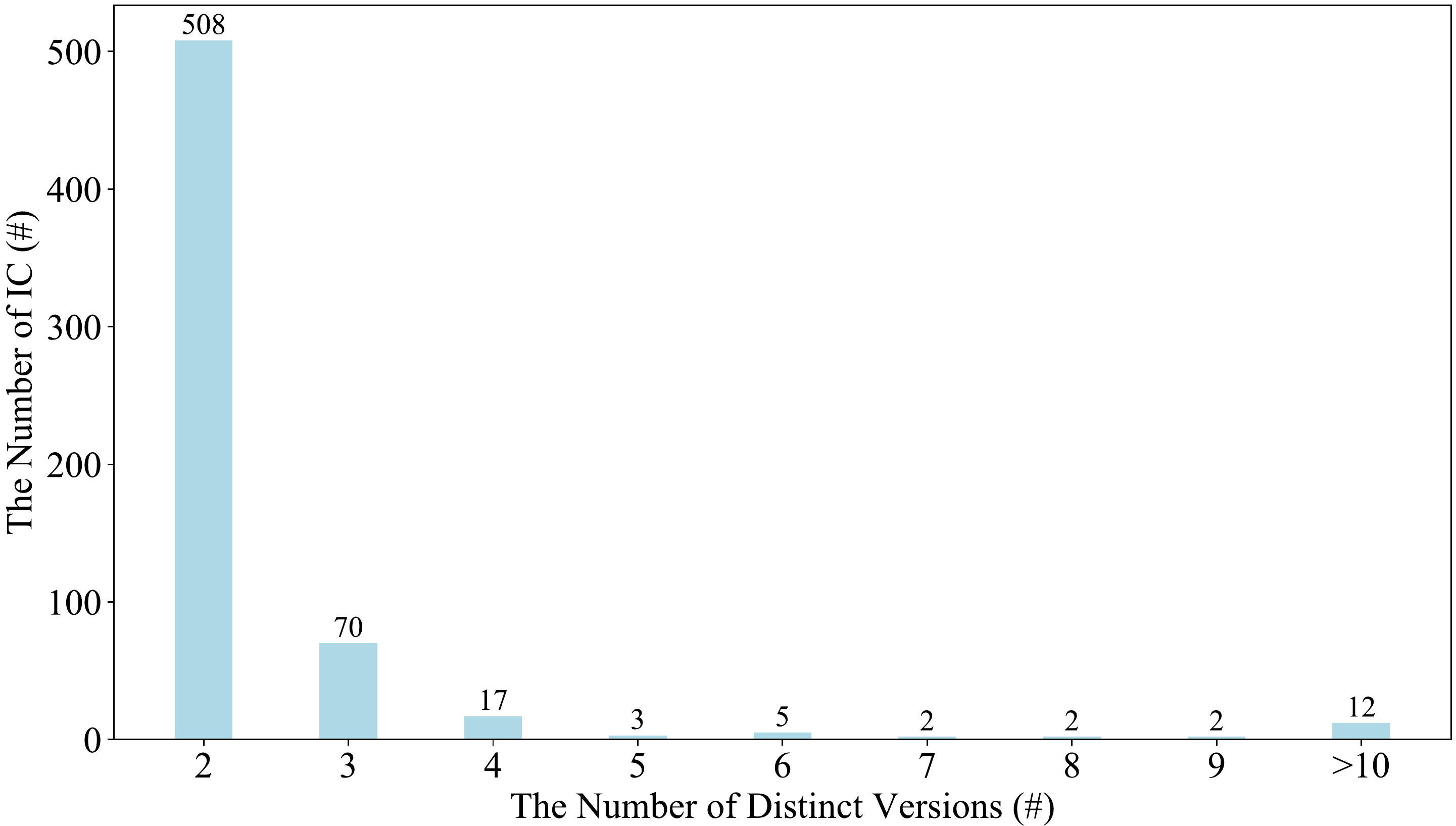}
    \vspace{0pt}
    \caption{Distinct Versions}
    \label{fig:rq2-distinct-version}
    \end{subfigure}
    ~~
    \begin{subfigure}[b]{0.49\textwidth}
    \centering
    \includegraphics[width=0.99\textwidth]{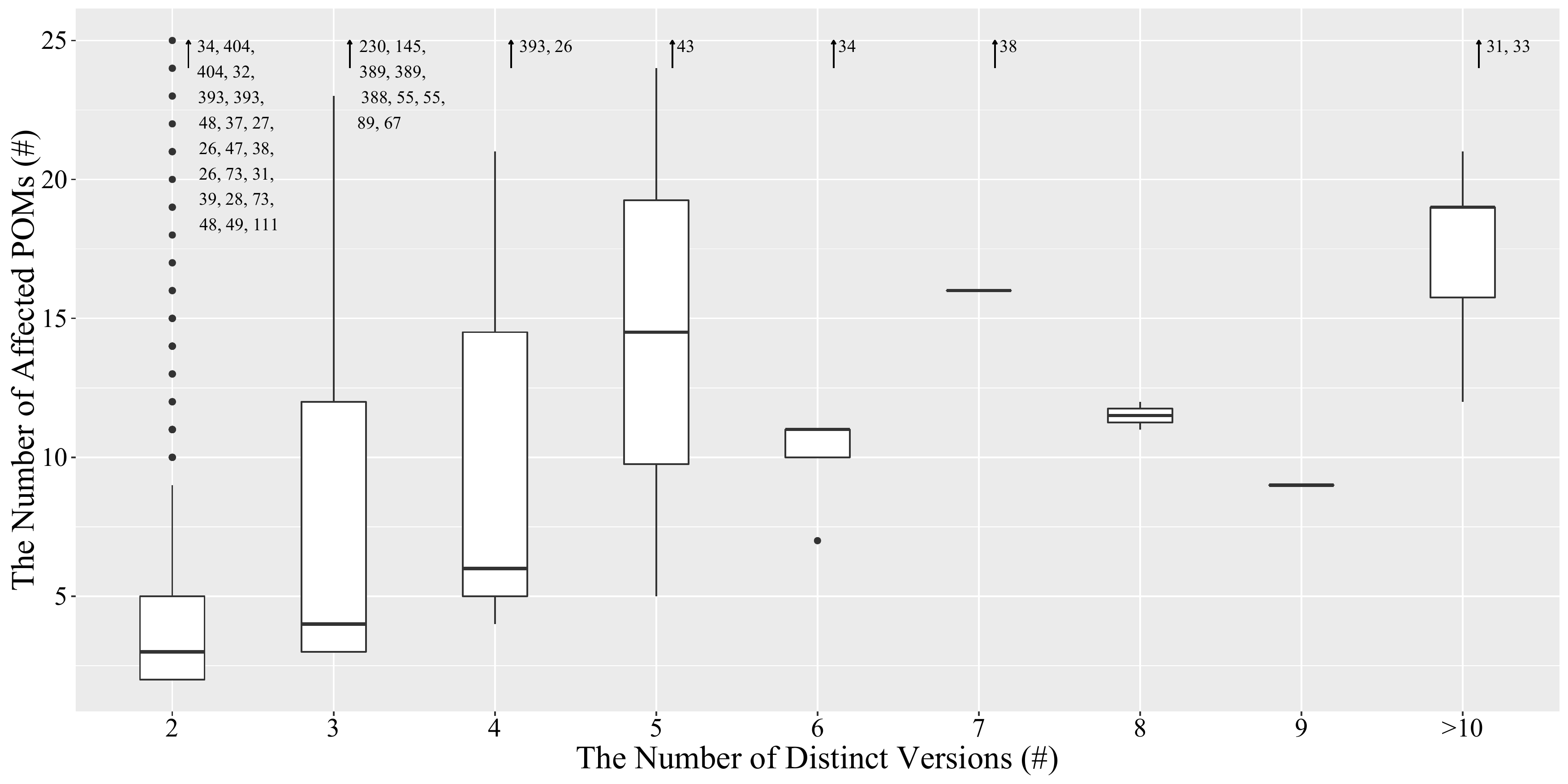}
    \vspace{0pt}
    \caption{Distinct Versions w.r.t. Affected POMs}
    \label{fig:rq2-distinct-affected}
    \end{subfigure}
    \vspace{-10pt}
    \caption{Distribution of Distinct Versions in IC}
    \label{fig:rq2-affected} 
\end{figure*}

\textbf{Fine-Grained Distribution.} The bars in Fig.~\ref{fig:rq1-distribution-poms} report the total number of projects whose number of POMs is in a specific range.~As we can see, nearly half (\todo{47.6\%}) of the projects have~less~than~10~POMs, and only \todo{22.3\%} of projects have more than 30 POMs. These results indicate that most projects have moderate complexity in modules. The four curves in Fig.~\ref{fig:rq1-distribution-poms} reports the distribution of IC, FC, TC and SL as projects' complexity in modules increases, while~the~four~curves in Fig.~\ref{fig:rq1-distribution-poms-affect} correspondingly present the ratio of projects that have~IC, FC, TC and SL. We can see~that,~the ratio of IC slightly increases and the ratio of projects having IC greatly increases. This indicates that as projects have more complexity in modules, library maintenance becomes more complicated, and hence there is a higher chance to introduce inconsistencies. Besides, the ratio of FC and TC does not decrease, and the ratio of projects having FC and TC even increases. This indicates that although projects become~more~complex~in~modules, developers may still willing to keep library~versions consistent. In that sense, \toolname can help developers systematically detect inconsistencies or false consistencies as early as possible.

%\begin{figure*}[!t]
%    \centering
%    \begin{subfigure}[b]{0.463\textwidth}
%    \centering
%    \includegraphics[width=0.99\textwidth]{fig/RQ1-iclibrank.pdf}
%    \caption{Inconsistent library}
%    \label{fig:rq1-iclibrank}
%    \end{subfigure}
%    \begin{subfigure}[b]{0.47\textwidth}
%    \centering
%    \includegraphics[width=0.99\textwidth]{fig/RQ1-fclibrank.pdf}
%    \caption{false consistent library}
%    \label{fig:rq1-fclibrank}
%    \end{subfigure}%
%    \vspace{-5pt}
%    \caption{Libraries that mostly Involved into IC and FC}
%\end{figure*}

%\textbf{Inconsistent Libraries.} Moreover, we rank libraries by their frequencies and select top 20 libraries in FC and IC(See Fig \ref{fig:rq2-ic} and \ref{fig:rq2-fc}). The top 20 libraries covers from logging libraries, unit testing libraries, IO utils to http/networking, which are popular libraries for code reuse. We find that com.google.guava:guava ranks first in inconsistent libraries, and it is not a surprising fact because to our knowledge, there already have been questions\cite{SO-21867295}\cite{SO-10918027} and issues\cite{HBASE-17908}\cite{JAVA-1328} about backward compatibility about it because guava is evolving rapidly. Also developers have to keep multiple versions to ensure the outdated code is executed well. In false consistent libraries, junit:junit ranks first and org.mockito:mockito-core ranks third, which are unit testing tools and logging library org.slf4j:slf4j-api ranks second. \todo{why junit harcoded most?}
 
 \begin{tcolorbox}[size=title, opacityfill=0.15]
\toolname detected \todo{621} library version inconsistencies and \todo{2,576} false consistencies, accounting for \todo{16.4\%} while affecting \todo{364} projects. As projects have more complexity in modules, it becomes more likely to introduce inconsistencies.
\end{tcolorbox}

\subsection{Severity Evaluation (RQ2)}\label{subsec:severity}

We analyzed the severity of a detected inconsistency or false~consistency (i.e., $\mathcal{M}_{lib}$) in terms of four indicators: 1) the number~of~POMs that are affected (i.e., $|\mathcal{M}_{lib}|$), 2) the ratio of POMs that are affected (i.e., $|\mathcal{M}_{lib}|$ / $|\mathcal{M}|$), 3) the number of distinct versions declared in $\mathcal{M}_{lib}$,  and 4) whether the versions of library dependencies in $\mathcal{M}_{lib}$ are all explicitly declared (i.e., hard-coded), all implicitly declared (i.e., via referencing a property), or declared~in~a~mixed~way.~The third indicator is only applicable for inconsistencies as false consistencies only have one version. The higher the first three indicators, the more versions are simultaneously adopted in more POMs, and thus the more severe the inconsistency~or false~consistency. For the fourth indicator, we regard explicit declaration is more severe than mixed declaration and implicit declaration because it indicates that developers seem to have no sense to harmonize library versions~via a property. We report the aggregated result over~all~consistencies or false consistencies for each of the four indicators.

\textbf{Affected POMs.} Fig.~\ref{fig:rq2-abs} presents the distribution of  IC and FC with respect to the number of affected POMs. \todo{67.8\%} of ICs and \todo{70.8\%} of FCs affect less than five POMs, and \todo{21.3\%} of ICs~and~\todo{12.9\%}~of~FCs affect more than ten POMs. On the other hand, Fig.~\ref{fig:rq2-rel}~reports~the~distribution of IC and FC with respect to the ratio of affected POMs. \todo{70.9\%} of ICs and \todo{53.5\%} of FCs affect less than 20\% of POMs, while \todo{9.7\%} of ICs and \todo{22.9\%} of FCs affect more than 50\% of POMs. These~results indicate that most ICs and FCs affect a relatively small number of POMs, but still around one-tenth of ICs and one-fifth of FCs~could involve a relatively large number of POMs.
 
\textbf{Distinct Versions.} Fig.~\ref{fig:rq2-distinct-version} reports the distribution of distinct~versions in inconsistencies. We can see that \todo{81.8\%} of ICs only have~two distinct versions, and only \todo{3.7\%} of ICs have more than five distinct version. Moreover, we generated a box plot for each bar in Fig.~\ref{fig:rq2-distinct-version} to measure the affected POMs. The result~is~reported in Fig.~\ref{fig:rq2-distinct-affected}, where the arrows indicate higher outliers that we hide to enhance the comprehension of the box plots. As the number of distinct versions in ICs increases, the number of affected POMs increases. In regard of the ICs that have two distinct versions, the median number of affected POMs is around \todo{three}. This indicates that most ICs are still manageable~if~developers want to harmonize them. Still, there are \todo{80} outliers in the first box plot, and some of them can affect around 400 POMs. We looked into these \todo{80} outliers, and found that in \todo{72 (90.0\%)} outliers, more than 80\% of the POMs use one version, while less than 20\% of the POMs use the other version. More interestingly, in \todo{58 (72.5\%)} outliers, one of the distinct version is only used~in~one POM. One potential reason is that developers have to use a specific version in a minority of POMs to avoid the heavy API backward incompatibility in them. Another potential reason is that developers are unaware of the minority of POMs that use a distinct version due to the complex POM inheritance graph.

 \begin{figure}[!t]
    \centering
    \begin{subfigure}[b]{0.22\textwidth}
    \centering
    \includegraphics[width=0.99\textwidth]{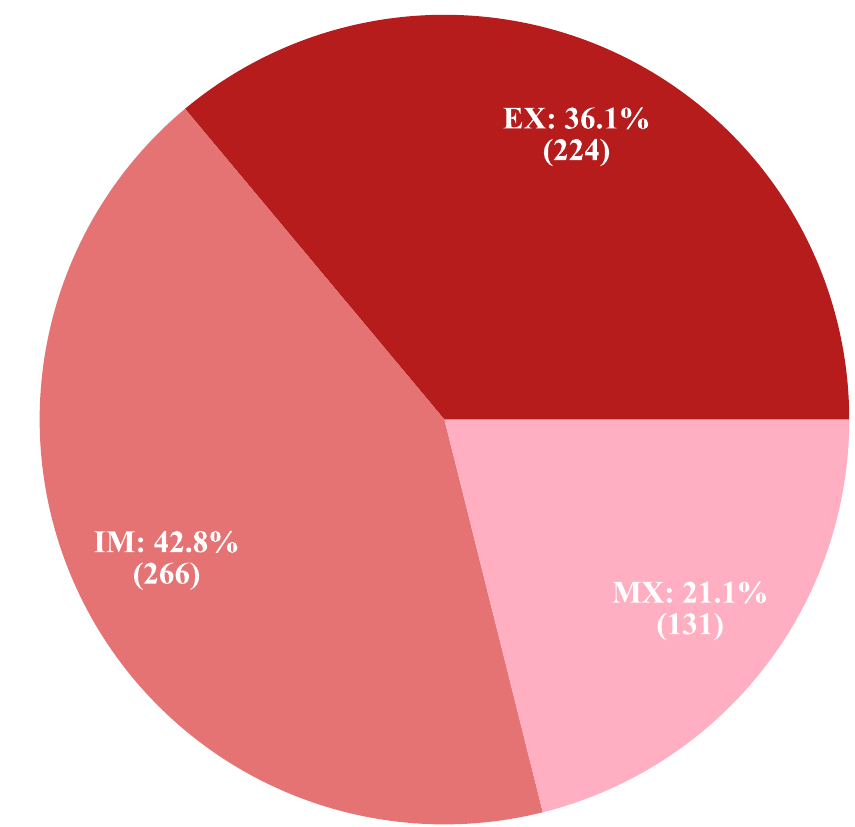}
    \vspace{-10pt}
    \caption{IC}
    \label{fig:rq2-ic-vd}
    \end{subfigure}
    ~
    \begin{subfigure}[b]{0.22\textwidth}
    \centering
    \includegraphics[width=0.99\textwidth]{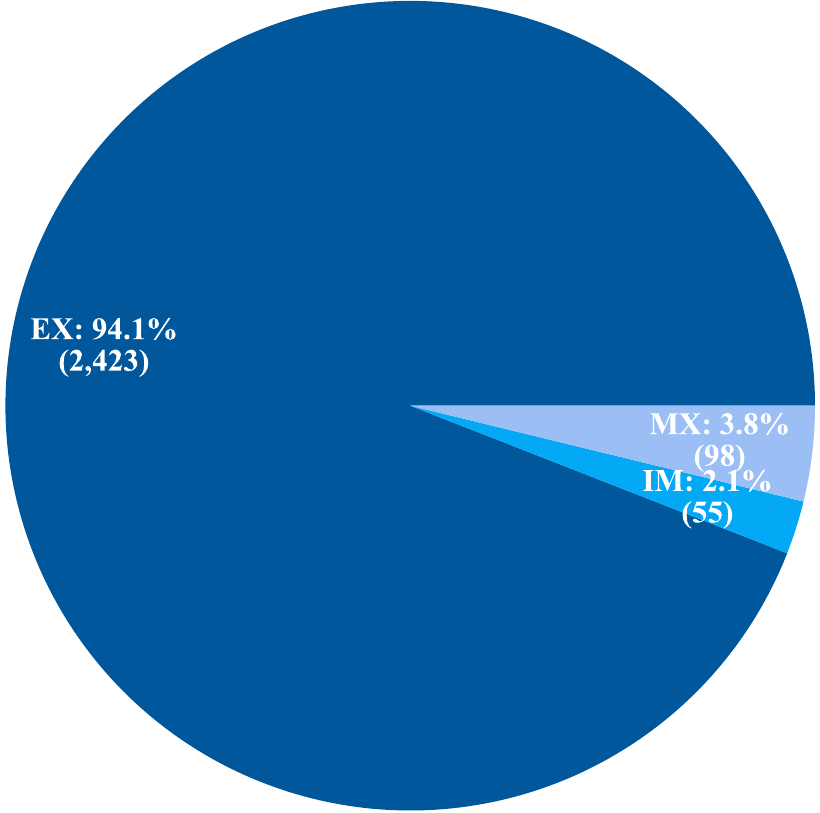}
    \vspace{-10pt}
    \caption{FC}
    \label{fig:rq2-fc-vd}
    \end{subfigure}
    \vspace{-10pt}
    \caption{Version Declaration Distribution in IC and FC}
    \label{fig:rq2-vd} 
\end{figure}

\textbf{Version Declaration.} Fig.~\ref{fig:rq2-vd} shows the distribution of version declarations (i.e., explicit declaration (EX), implicit declaration (IM) and mixed declaration (MX)) for IC and FC. It turns out that  \todo{94.1\%} of FCs declare versions by hard-coding. This means that developers need to change all the affected POMs at the same time to keep these FCs consistent rather than turning such FCs into ICs,~which~is~actually a huge but avoidable maintenance cost. Besides, \todo{36.1\%}~of~ICs declare versions by hard-coding. This shows that hard-coding~version numbers is probably not a good practice, and it tends to introduce inconsistencies. \todo{63.9\%}~of~ICs~include implicitly~declared~versions.~This means that developers~already~have~the~sense~to declare versions~by referencing a property for reducing library maintenance cost,~but~IC still exists. As revealed by our survey (see~Sec.\ref{sec:survey}), there~are multiple reasons for not harmonizing inconsistencies intentionally. We~attempted to manually look into these cases~to~determine whether~our detected inconsistencies were intentionally~kept.~However,~it~is~very challenging to confirm the underlying~reasons as they~are~often~business logic-related. This is also one of the reasons that we allow~developers to interact with \toolname. Nevertheless, we did find~some cases that conformed to our survey; but we also found one case that was not reported by our survey, i.e., developers intentionally~declare multiple properties for different versions of the same~library~to~provide comprehensive support for different runtime environments. For example, project \texttt{memcached-session-manager} is a tomcat session manager that keeps sessions in memcached or Redis, for highly available, scalable and fault tolerant web applications. It declares four properties for version 6.0.45, 7.0.85, 8.5.29 and 9.0.6 for various tomcat dependencies, as it is currently working with tomcat 6.x, 7.x, 8.x and 9.x (as explained in its \texttt{README} file).

%For example, in project \texttt{apache} \texttt{camel}, three versions (i.e., 18.0, 19.0 and 20.0) of \texttt{guava} are adopted, and most modules rely on 18.0, while only two modules update to 19.0 and 20.0。 One of the reason is that different modules depends on different versions and it requires certain cost to avoid backward compatibility issues so that version harmonization moves down on the priority list. 

\begin{tcolorbox}[size=title, opacityfill=0.15]
\todo{67.8\%} of ICs and \todo{70.8\%} of FCs affect less than five POMs. \todo{81.8\%} of ICs only have~two distinct versions, affecting a median~number of \todo{three} POMs. \todo{36.1\%}~of~ICs and \todo{94.1\%} of FCs declare all versions by hard-coding. Overall, the severity of ICs and FCs is relatively not high. 
\end{tcolorbox}

\subsection{Efforts Evaluation (RQ3)}\label{subsec:effort}

\begin{figure}[!t]
    \centering
    \includegraphics[scale=0.24]{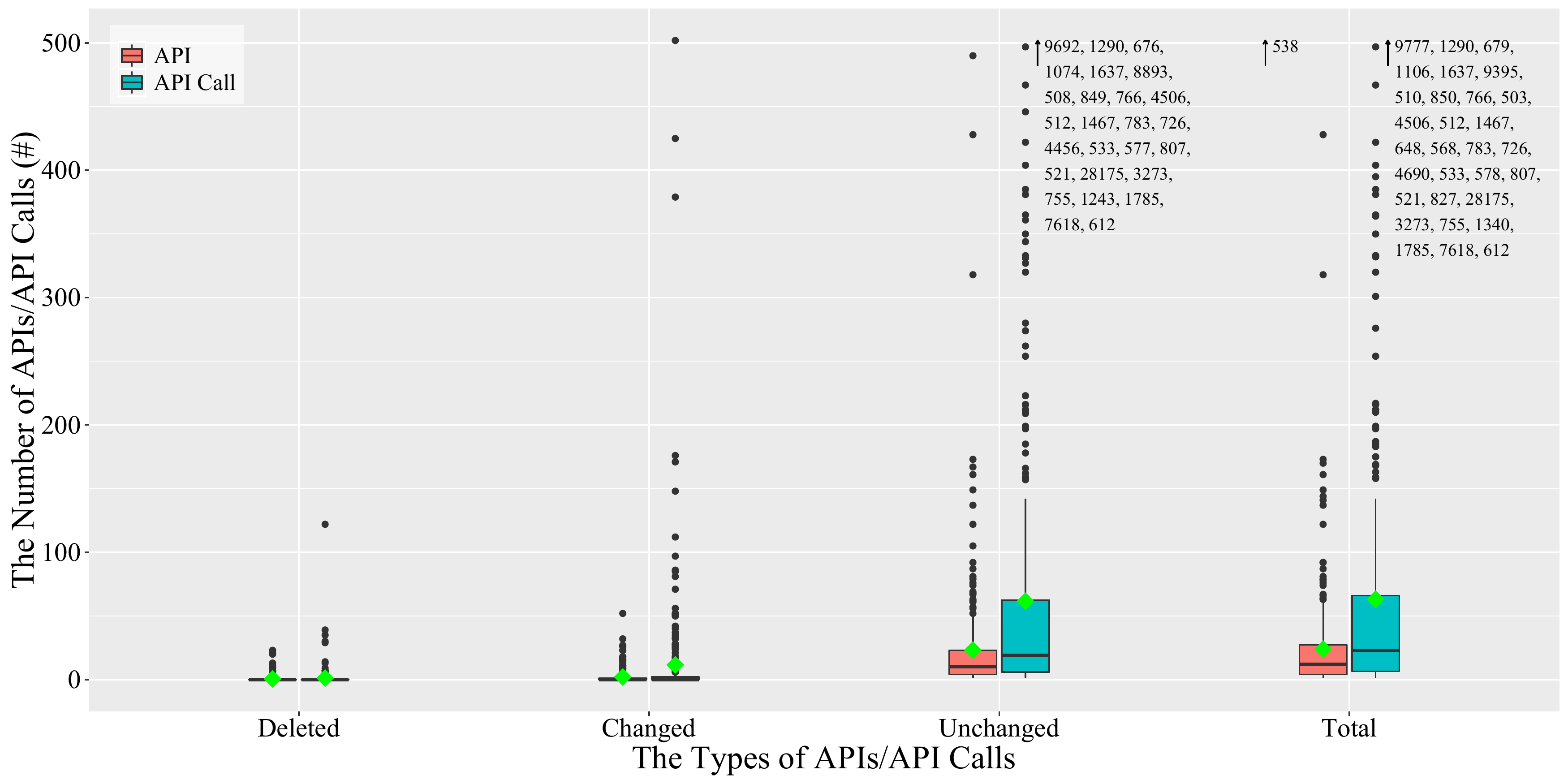}
    \vspace{-15pt}
    \caption{Harmonization Efforts}
    \label{fig:rq3-effort} 
\end{figure}

We analyzed the harmonization efforts for each of the \todo{621} inconsistencies for each candidate harmonized version. We report~the~results for the candidate version with the least harmonization efforts selected based on our default ranking (see Sec.~\ref{subsec:suggest}).~Fig.~\ref{fig:rq3-effort}~shows~two box plots (one denotes the number of APIs,  and the other denotes the number of API calls) for deleted, changed, unchanged, and total library APIs that are called. Overall, \todo{190} (\todo{30.6\%}) ICs have~no~harmonization efforts; i.e., all the invoked library APIs are not changed~in the suggested harmonized version. In the remaining \todo{431} ICs,~on~average, \todo{1} and \todo{2} of the \todo{24} called library APIs are respectively deleted and changed in the suggested harmonized version, affecting \todo{1}~and~\todo{12} of the \todo{63} library API calls, as~indicated by the green diamonds. These results indicate that the harmonization efforts with respect to the number of deleted and changed APIs seem small. However,~the~actual harmonization efforts depend on how the deleted or changed APIs affect the business logic. Therefore, we choose to provide~developers with~detailed API-level report to assist them in determining where and whether to harmonize.

%average 1,  1, 2, 12, 24, 63
%median 0, 0, 0, 0, 12, 23

\textbf{Replacement API to Deleted APIs.} In these suggested harmonized versions for the \todo{621} inconsistencies, a total of \todo{1,798} library APIs are deleted. Our documentation-based replacement API suggestion method (see Sec.~\ref{subsec:refactor}) successfully find the replacement APIs for \todo{207 (11.5\%)} of them. Such a low coverage indicates the potential efforts in refactoring source code. Contrarily, our survey indicates that automatic source code is a less useful feature. This~is~also~why we decide to design such an API suggestion method that only suggests replacement APIs with 100\% accuracy.

\textbf{Case Studies.} Here we choose a popular project \texttt{Apache} \texttt{Tika}~to demonstrate what our report tells to the developers. \texttt{Apache} \texttt{Tika}~is a toolkit for detecting and extracting metadata and structured text content from various documents using existing parser libraries. It is a project of the Apache Software Foundation. \toolname identifies four inconsistencies and six false consistencies.

One of the inconsistencies is about library \texttt{commons-cli}. It involves three modules: \texttt{tika-server}, \texttt{tika-batch} and \texttt{tika-eval}. The first module explicitly declares \texttt{commons-cli} with version~1.2. The latter two reference two properties with the same property~name \texttt{cli.version} defined in their own POM file, and both of them~declare version 1.4. Therefore, an inconsistency occurs. Version~1.4~is suggested as the harmonized version as it leads to the smallest~number of calls to the deleted and changed library APIs, indicating that updating \texttt{commons-cli} from 1.2 to 1.4 in \texttt{tika-server} solves the~inconsistency. It turns out that the number of called library APIs in \texttt{tika-server} is 5, and there are totally 33 library API calls.~These 5 library APIs are all changed in version 1.4. Finally, a lowest~common ancestor POM file, \texttt{tika-parent/pom.xml}, is located, where~a new property specifying version 1.4 is declared. The two old properties declared separately in \texttt{tika-batch} and \texttt{tika-eval} can be removed safely because no other library~dependency references~them.~\todo{In~fact, according to the comment in the POM file, developers actually have already noticed this inconsistency, and thought about moving the property declaration into \texttt{tika-parent/pom.xml} (as suggested by \toolname) to harmonize the inconsistency. However, the harmonization was postponed due to migration efforts.}

One of the false consistencies is about \texttt{json-simple}. It involves two modules: \texttt{tika-parsers} and \texttt{tika-translate}. In these two~modules, \texttt{json-simple} is declared explicitly with version 1.1.1. Similar to the previous inconsistency example,  \toolname searches~the~POM inheritance graph and locates \texttt{tika-parent/pom.xml} as a common parent POM file; then it declares a property of version 1.1.1, and refactors \texttt{json-simple} dependency declaration in \texttt{tika-parsers} and \texttt{tika-translate} to implicitly reference the new property. 

 \begin{tcolorbox}[size=title, opacityfill=0.15]
 In \todo{190} (\todo{30.6\%}) ICs, all the called library APIs are not changed~in the suggested harmonized version. In the remaining \todo{431} ICs,
 
 on~average, \todo{1} and \todo{2} of the \todo{24} called library APIs are deleted~and changed, affecting \todo{1}~and~\todo{12} library API calls. Overall,~the~harmonization efforts seem relatively small but the true efforts are often application-specific.
\end{tcolorbox}

\subsection{Developer Feedback (RQ4)}\label{subsec:feedback}
   
To understand developers' feedback about \toolname, we targeted \todo{621} inconsistencies in \todo{152} projects, and sent our generated report to the developers of these projects. Within one week, \todo{16} developers replied. \todo{8} of them explicitly commented that version inconsistency is certainly a problem for library maintainers,~and~our tool~and~report are useful; e.g., ``\textit{the problem you're describing is very real, and I have encountered it myself in my day-to-day job several times}'', ``\textit{keep up the good work with your harmonization tool. It definitely sounds interesting!}'', and ``\textit{the cool reports here helped me~find one real issue, thanks!}''. \todo{5} of them did not comment on our tool, but only~discussed the inconsistencies, and \todo{3} of them were no longer Java developers or no longer in charge of the projects.

Moreover, \todo{4} developers confirmed the inconsistencies but explained that they were intentionally kept due to API compatibility, and \todo{1} developers confirmed and quickly fixed the consistency. As~we crawled the project repositories several months before our report, \todo{4} developers asked us to re-generate the report for their current repositories, and we are still waiting for their further feedback.~The others are still under discussion.

Interestingly, a developer~from \texttt{hadoop} confirmed that adopting consistent libraries is one of their common practice, but ``\textit{people~neglect to do this; when that's found we will~pull the explicit version~declaration out and reference from hadoop-project; adding the import~there if not already present. Therefore any duplicate declaration of a dependency with its own <version> field in any module other than hadoop-project is an error. Your dependency graphs are helpful here}''. It is also worth mentioning that \todo{4} developers commented~that they also cared about inconsistencies in transitive dependencies,~but also said that ``\textit{it is also very hard to fix, since the source code is not owned by me}''. This is why we only focus on direct dependencies.

\begin{tcolorbox}[size=title, opacityfill=0.15]
Half of the responded developers thought that our tool and report are useful. \todo{5} inconsistencies have been confirmed and \todo{1} of them has been harmonized.
\end{tcolorbox}

\subsection{Discussions}

\textbf{Threats to Survey.} First, we chose an online survey with GitHub developers instead of a face-to-face interview study with industrial developers, because it is difficult to recruit industrial developers~for interviews at a reasonable cost, and an online study~allows~us~to~recruit a relatively large number of developers. Second, we decided~to not offer compensation but kindly ask participants to voluntarily take the survey. As a result, we expected that GitHub developers who were really interested in library version inconsistencies and well motivated would participate in this survey. This instead could improve the quality of our survey to avoid potential cases~that~participants only wanted the compensation but answered haphazardly. %Third, like in any online survey, some participants may still answer the survey haphazardly. To avoid such low-quality data, we actually removed any obviously low-quality data (e.g., responses that are entirely invective) before analysis, but we cannot discriminate perfectly.

\textbf{Threats to Evaluation.} First, as we have not integrated our~tool into the build process, it is not feasible for us to empirically evaluate the soundness of our tool in refactoring POMs on a large-scale of diverse projects. However, we refactor POMs in such a non-invasive way that does not change the inheritance relationship among POMs. We believe our POM refactoring is sound. Second, due to the same reason, we generated reports about inconsistencies and sent reports to relevant developers for obtaining their feedback instead of letting developers directly use our tool on their~projects.~While~this~may~not get the first-hand information from developers, it relieves the burden of developers to install our tool and only focuses on the results. We believe this can help us obtain more feedback.

%Secondly, once the versions are harmonized into one, the compatibility issue is still not fully confirmed. And compatibility issue is a huge problem when migrating. Although we try to offer API replacements on the basis on JavaDocs, how to replace the rest of the deleted APIs still remains a manual migration job to developers which need comprehensive testing jobs.

\textbf{Limitations.} First, due to the well-known limitation of static~analysis, our generated API call graphs could be unsound (e.g., due to reflection), which~will~affect the~precision of our API-level harmonization efforts analysis. We will investigate a combination of static analysis and dynamic analysis to make the call graph generation more~precise. Second, we currently only target Maven Java projects, but~there are several other automated build tools such as Gradle and Ant. The reasons we currently choose Maven are~that~1)~Maven~is the most widely-used build tool for Java project, and~2)~many Maven projects have been developed for decades and these long-history projects may be mostly beneficial from our tool. Given the positive feedback from developers, we plan to develop corresponding tools for other automated build tools.

% !TeX root = ../main.tex

\section{Related Work}

In this section, we review the related work in four areas:~library~analysis, API evoluation, API adaptation, and library empirical studies.

\subsection{Library Analysis}

Patra et al.~\cite{patra2018conflictjs} analyzed JavaScript library conflicts caused by the lack of namespaces in JavaScript, and proposed ConflictJS to first use dynamic analysis to identify potential conflicts and then use targeted test synthesis to validate them.~Wang~et~al.~\cite{Wang2018DCM} investigated the manifestation and fixing patterns of dependency conflicts in Java, and designed \textsc{Decca} to detect dependency conflicts and assess their severity via static analysis. Wang et al.~\cite{Wang2019CIH}~also~proposed \textsc{Riddle} to generate crashing stack traces for detected dependency conflicts. Such dependency conflicts are one of the bad~consequences of inconsistent library versions.

Cadariu et al.~\cite{cadariu2015tracking} proposed an alerting tool to notify developers about Java~library dependencies with security vulnerabilities. Mirhosseini and Parnin~\cite{mirhosseini2017can} compared the usage of pull requests and badges to notify outdated npm packages. These approaches~only detect the inclusion of vulnerable libraries. To further determine~if the vulnerable library code is in the execution path of a project,~Plate et al.~\cite{plate2015impact} applied dynamic analysis to check whether the vulnerable methods were executed by a project; and Ponta et al.~\cite{PontaPS2018beyond}~extended~it by combining dynamic analysis with static analysis. It is interesting to also consider security vulnerabilities as another factor~when~we recommend harmonized versions in \toolname.

Bloemen et al.~\cite{bloemen2014gentoo} analyzed the evolution of the Gentoo package dependency graph, while Kikas et al.~\cite{kikas2017structure} and Decan et al.~\cite{decan2019empirical} compared the evolution of dependency graphs in different ecosystems. Kikas et al.~\cite{kikas2017structure} and Decan et al.~\cite{Decan2018ISV} also investigated the impact of security vulnerabilities on the dependency graph. Zimmermannet et al.~\cite{Zimmermann2019small} further modeled maintainers and vulnerabilities into~the dependency~graph in the npm ecosystem, and systematically analyzed the risk of attacked packages and maintainers and vulnerabilities. \toolname can be extended to support library version inconsistency analysis on the ecosystem-level dependency graph. 

To the best of our knowledge, no previous work has systematically investigate library version inconsistency.

\subsection{API Evolution}

A large body of studies have been focused~on~API evolution to analyze how developers react to API evolution~\cite{robbes2012developers,sawant2016reaction,hora2015developers,McDonnell2013ESA,Bogart2016BAC}, how APIs are changed and used~\cite{wu2016exploratory}, how API stability~is measured~\cite{raemaekers2012measuring}, how API stability affects Android apps'~success~\cite{Linares-Vasquez2013ACF}, how refactoring influences API breaking~\cite{Dig2006AES,Kim2011EIR,Kula2018ESI}, how and why developers break APIs~\cite{jezek2015java,xavier2017we,brito2018and}, how API breaking impacts client programs~\cite{xavier2017historical,raemaekers2017semantic}, etc. Moreover, several advances have been made to detect API breaking. Previous work mostly uses theorem proving \cite{McCamant2003PPC,mccamant2004early,lahiri2012symdiff,godlin2013regression,felsing2014automating} or symbolic execution \cite{trostanetski2017modular,Mora2018CEC}, but has scalability issues when detecting breaking APIs in real-life program. Recently, testing techniques have been used to detect breaking APIs. Gyori et al.~\cite{gyori2018evaluating} relied on regression tests, while Soares et al.~\cite{soares2010making} generated new tests to detect behavior changes in refactored APIs. Mezzetti et al.~\cite{mezzetti2018type} and M{\o}ller and Torp~\cite{moller2019model} targeted Node.js libraries, and used model-based testing to detect type-related breaking (i.e., changes~to API signatures). Similarly, Brito et al.~\cite{brito2018apidiff} used heuristics to statically detect type-related changes in Java libraries. However, it is an open problem to detect behavior changes when API signatures~are not changed but the API bodies are changed. We will extend such approaches to improve the precision of our effort analysis.

\subsection{API Adaptation}

A large number of advances have been made to adapt client programs to API evolution according to change rules. Change rules can be manually written by developers~\cite{Chow1996SUA,Balaban2005RSC},~automatically recorded from developers~\cite{henkel2005catchup}, derived through~API~similarity matching~\cite{xing2007api}, mined from API usage changes in libraries~themselves~\cite{dagenais2009semdiff,dagenais2011recommending} as well as client programs~\cite{schafer2008mining,Nguyen2010GAA,fazzini2019automated}, and extracted~by a combination of some of these methods~\cite{wu2010aura}. Several empirical~studies have also been proposed to investigate the effectiveness of these methods~\cite{cossette2012seeking, wu2015impact}. They found that these methods only achieved~an average accuracy of 20\%, but still could help developers. Currently, we only recommend change rules that are extracted from documentation and hence are absolutely correct, and we will integrate~these methods to recommend undocumented change rules.

%Mining migration rule in API level could offer support when undergoing a library migration. Dagenais et. al. proposed \textit{SemDiff}\cite{dagenais2009semdiff}\cite{dagenais2011recommending}, a tool that recommends replaces for framework methods that were accessed by a client program and deleted during the evolution of the framework. Schafer et. al.\cite{schafer2008mining} proposed to mine framework usage change rules from already ported instantiations of two versions of a framework. Similar technique is used for class library migration on the basis of type constraints\cite{Balaban2005RSC}. Aside from mining change rules, text similarity\cite{wu2010aura} and graph-based\cite{Nguyen2010GAA} approach have integrated for better result. Wu et. al.\cite{wu2010aura} applies combination of call dependency and text similarity to automatically identify change rules for one-replaced-by-many and  many-replaced-by-one. Besides, Dig et al.\cite{dig2008reba} proposes to a technique to generate compatibility layers to mitigate the adaptation problem.

\subsection{Library Empirical Studies}

 A large body of studies has been~focused on characterizing the usage and update practice of libraries in different ecosystems, e.g., the usage trend and popularity~of~libraries and APIs~\cite{bauer2012structured,bauer2012understanding,mileva2009mining,kula2017exploratory,Mileva2010,hora2015apiwave,lammel2011large,de2013multi,qiu2016understanding,wittern2016look,abdalkareem2017developers,li2016investigation},~the~practice~of~updating library versions~\cite{kula2018developers,bavota2013evolution}, the~latency~of~updating~library versions~\cite{kula2015trusting,cox2015measuring,lauinger2017thou,Decan2018evolution}, and the reason of updating~or~not~updating~library versions~\cite{bavota2013evolution,bavota2015apache,Derr2017KMU,kula2018developers}. To the best of our knowledge, we are the first to systematically understand library version inconsistency in real-life projects through a survey with GitHub developers.

% !TeX root = ../main.tex

\section{Conclusions}\label{sec:conclusion}

In this paper, we have conducted a survey with \todo{131} Java developers from GitHub to collect the first-hand information about root~causes, detection methods, reasons for fixing or not fixing, fixing strategies, fixing~efforts, and tool expectations on library version inconsistencies. Our~survey suggests several insights, e.g., tools~are~needed to proactively~locate~and harmonize inconsistent library versions, and such tools need to interact with developers and provide API-level harmonization efforts. Based on such insights, we have developed \toolname, the first interactive, effort-aware technique~to~harmonize inconsistent library versions in Java Maven projects. We have evaluated \toolname against \todo{443} Java~Maven projects from GitHub. \toolname successfully detects \todo{621} library version inconsistencies, covering \todo{152} of projects, as well as \todo{2,576}~false consistencies, covering \todo{346} projects.  The average harmonization efforts~are that \todo{1} and \todo{2} of the \todo{24} called library APIs are respectively deleted and changed in the harmonized version, affecting \todo{1} and \todo{12} library API calls. Moreover, \todo{5} library version inconsistencies have been confirmed, and \todo{1} of them has been harmonized by developers. In future, we plan to integrate \toolname into the build process so that developers can more intensively and naturally interact with \toolname. We also hope to extend \toolname to support other automation build tools (e.g., Gradle), and develop advanced API breaking analysis techniques to improve the accuracy of our API-level effort analysis.

%\begin{acks}
%\end{acks}

\bibliographystyle{ACM-Reference-Format}
\bibliography{src/reference}

\end{document}